\documentclass[reprint,aps,pre,superscriptaddress]{revtex4-1}
\usepackage{amsmath}
\usepackage{amsthm}
\usepackage{amsfonts}
\usepackage{bm}
\usepackage{enumitem}
\usepackage[version=4]{mhchem}
\usepackage{chemformula}
\usepackage{graphicx}
\usepackage{caption, siunitx}

\renewcommand{\epsilon}{\varepsilon}
\newcommand{\change}[1]{\textcolor{black}{#1}}

\font\tiny=cmr5

\def\wrapsign#1#2#3{\wrap{{\textcolor{#2}{#1}}}\lower 4pt\hbox{\tiny #3}}
\def\wrap#1{[\kern-2pt[\thinspace{#1}\thinspace]\kern-2pt]}

\begin{document}

\setchemformula{kroeger-vink=true}

\title{ Theory of layered-oxide cathode degradation in Li-ion batteries by oxidation-induced cation disorder }

\author{Debbie Zhuang}
\affiliation{Department of Chemical Engineering, Massachusetts Institute of Technology \\ 77 Massachusetts Avenue, Cambridge, MA 02139}
\author{Martin Z. Bazant}
\email{bazant@mit.edu}
\affiliation{Department of Chemical Engineering, Massachusetts Institute of Technology \\ 77 Massachusetts Avenue, Cambridge, MA 02139}
\affiliation{Department of Mathematics, Massachusetts Institute of Technology \\ 77 Massachusetts Avenue, Cambridge, MA 02139}
\date{\today}

\begin{abstract}
    
 Disorder-driven degradation phenomena, such as structural phase transformations and surface reconstructions, can significantly reduce the lifetime of Li-ion batteries, especially those with nickel-rich layered-oxide cathodes.
    We develop a general free energy model for layered-oxide ion-intercalation materials as a function of the degree of disorder, which represents the density of defects in the host crystal. The model accounts for defect core energies, long-range dipolar electrostatic forces, and configurational entropy of the solid solution.
    In the case of nickel-rich oxides, we hypothesize that nickel with a high concentration of defects is driven into the bulk by electrostatic forces as oxidation reactions at the solid-electrolyte interface \change{ reduce nickel and either evolve oxygen gas or oxidize the organic electrolyte} at high potentials ($>4.4V$ vs. Li/Li$^+$). The model is used in battery cycling simulations to describe the extent of cathode degradation when using different voltage cutoffs, in agreement with experimental observations that lower-voltage cycling can substantially reduce cathode degradation. The theory provides a framework to guide the development of cathode compositions, coatings and electrolytes to enhance rate capability and enhance battery lifetime. The general theory of cation-disorder formation may also find applications in electrochemical water treatment and ion separations, such as lithium extraction from brines, based on competitive ion intercalation in battery materials.
    
\end{abstract}

\maketitle

\section{Introduction}

\subsection{Motivation}

As Li-ion batteries continue to revolutionize energy storage and power global electrification, it is increasingly important to understand the microscopic degradation mechanisms that limit their efficiency, rate capability, and lifetime.
Degradation is exacerbated by efforts to increase energy density, while lowering material costs. This tradeoff is well illustrated by nickel-rich cathode materials, based on nickel-manganese-cobalt (NMC) layered oxides \cite{chakraborty2020layered, whittingham2004lithium}: as scarce and expensive cobalt is replaced by more plentiful and affordable nickel, the nickel-rich oxide cathode degrades more easily at the high voltages required for high energy density batteries.
The microscopic mechanisms are still poorly understood, so it is critical to develop a predictive theory of degradation, in order to understand and optimize this tradeoff.

Nickel-rich materials degrade with cycling by a variety of possible mechanisms \cite{kabir2017degradation, jiang2021review, li2020degradation, noh2013comparison}, such as phase transformations, cation disorder, surface reconstruction \cite{lin2014surface, yan2015evolution}, particle cracking \cite{trevisanello2021polycrystalline, heenan2020identifying, mao2019high}, and transition-metal dissolution \cite{evertz2016unraveling, wandt2016transition}.
Notably, the various possible degradation mechanisms depend on the specific transition-metal chemical properties of different nickel-rich materials.
Some nickel-rich degradation mechanisms, such as transition metal dissolution, can affect degradation at the anode through increasing the conductivity of the solid-electrolyte interphase on the graphite anode.
This happens by the incorporation of transition metal ions dissolved in the solution from cathode degradation \cite{zhan2018dissolution, zheng2012correlation}.
An additional complexity is that nickel-rich layered oxides are currently being developed into many different compositions with added transition metals such as cobalt, manganese, and aluminum, which all have varying chemical properties.
The degradation mechanisms of nickel-rich materials all are coupled, with cation disorder driving much of the bulk phase transitions and the surface phase transformations \cite{pender2020electrode}. 
Here we seek to elucidate the physical mechanisms behind degradation of nickel-rich cathode materials to increase battery lifetime and reduce safety risks during operation from degradation-induced short circuits.

Although Li-ion battery capacity fade has been the focus of extensive research, most studies have focused on degradation mechanisms in the standard graphite anode, such as solid electrolyte interphase growth \cite{pinson2012theory, attia2019electrochemical,das2019electrochemical, von2020solid} and lithium plating \cite{gao2021interplay, finegan2020spatial}. 
However, recent experiments have shown that the amount of cathode and anode degradation in batteries is on par with each other, especially in nickel-rich materials \cite{sabet2020non}.
Cycling experiments with microscopy techniques have been performed to visualize cathode degradation at the atomic scale \cite{lin2014surface, maleki2019controllable, yan2015evolution, yan2017atomic}, providing necessary experimental support for modeling at the particle level.
However, there have been few attempts to model cathode degradation beyond atomistic simulations or machine learning with limited physical insight \cite{qian2021understanding}.
Atomic scale studies using density functional theory and other methods have been used to study cathode degradation \cite{lee2012atomistic, das2017first, xiao2019understanding} to calculate the formation energies of these phases and the relationship between diffusivity coefficients and defects, but because of the scale at which degradation happens in a battery, this is impossible to translate directly into a porous electrode scale battery model.
The timescales of interest in atomic scale modeling are on the scale of nanoseconds at most, but the scale at which a battery operates is from hours to months, especially when degradation starts becoming of interest.
There is a crucial need to develop cathode degradation models that can be applied at the porous electrode scale in battery simulations.

Phase transformations at the surface and bulk, one of the main degradation mechanisms in nickel rich materials, have been observed since the initial characterization of nickel rich materials \cite{li1992crystal, das2017first, ceder1999phase, rougier1996optimization}.
Phase transformations and surface reconstruction, shown in Fig. \ref{fig:fig1}a, increase after cycling \cite{yan2017atomic}, and the phases formed are highly dependent on the material used \cite{das2017first, xiao2019understanding}.
Based on the varying cycling protocols used during (de)intercalation, the amount of phase transformations and the thicknesses of the surface reconstruction layers change \cite{yan2017atomic, marker2019evolution}.
Spinel, rock salt, and disordered rock salt phases, along with other phases such as $\gamma$ phases have all been measured experimentally \cite{das2017first, yan2017atomic, lin2014surface}, but no agreements have been reached over the phases formed, except that the phases are denser than the original layered phases.
These dense surface phases affect operation of a battery, modifying the kinetics and transport of these materials, causing batteries to become unusable after a certain point in their cycle life \cite{sallis2016surface, xu2021bulk, dixit2017origin}.

These phase transformations in the bulk and surface of cathode materials are a well-known degradation mechanism in nickel rich materials that have been studied with many experimental imaging techniques \cite{lin2014surface, zhu2020atomic, ko2020degradation}.
Computationally, density functional theory and Monte Carlo simulations have been used to study the effect of cation disorder on these phases \cite{das2017first, lee2014unlocking, xiao2019understanding, koyama2004systematic}.
Limitations based on computational power create difficulties in modeling the entire disorder process using quantum mechanical first-principles methods.
Understanding and reducing the degradation of batteries is an important barrier to the continued electrification of our current energy storage systems \cite{kabir2017degradation, li1992crystal}.
Thus, studying the amount of disorder, which drives the transitions to denser surface phases, is critical to studying the long-time operation of batteries and the continued expansion of decarbonized energy storage methods \cite{schiffer2017electrification}.

Here, we propose a physics-based mesoscale model that can be used to predict the long-term effects of cation disorder and phase transformations of nickel rich materials on degradation.
The free energy is based on electrostatics in periodic layered crystal structures and does not require any empirical fitting of data. It is derived directly from measured material properties, such as electronegativities of the transition metals.
In addition, this model can be applied to any form of layered nickel-rich battery material to study the effects of disordered regions.   
The thermodynamic model is then coupled to a simple model of the irreversible surface oxidation reaction (with a fitted rate constant and onset potential) and cation diffusion into the bulk, in order to predict the dynamics of degradation in layered oxides.

\subsection{Background}

It is important to note that the phase transformations associated with surface degradation are driven by defects, which can be created in the synthesis of these materials \cite{wang2020tuning, chen2011study} and also increase during battery operation \cite{yabuuchi2007changes}.
We take advantage of this to aid our modeling of degradation in nickel-rich cathodes, as well as go through some of the history in physics of using lattice models and dipoles to model structures.
Defects trigger phase transformations to denser phases, such as rock salt, spinel, or disordered rock salt structures usually found at the surface of NMC materials \cite{yan2015evolution, yan2017atomic, xiao2019understanding, maleki2019controllable, ji2019hidden}.

The main defect for nickel rich crystal structures is the antisite defect \cite{lee2012atomistic}, which can be observed as a kind of Schottky defect in an ionic solid lattice  \cite{shewmon2016diffusion, mehrer2007diffusion, kittel1996introduction}.
The description of the ``anion'' defect in our case is not an ionic defect but a negatively charged electron, while the cation defect is a lithium ion in the analogue to Schottky disorder.
The equilibrium concentration of Frenkel/Schottky defects is usually denoted by equilibrium constants using the law of mass action, $K_{eq} = e^{-G/k_B T}$, in terms of the free energy of formation $G$ \cite{kittel1996introduction}.
However, it is challenging to estimate formation energies without atomic scale calculations.
We instead turn to a method driven by topological defects in physics of studying this kind of disorder.

\begin{figure*}[t]
\includegraphics[width=\textwidth]{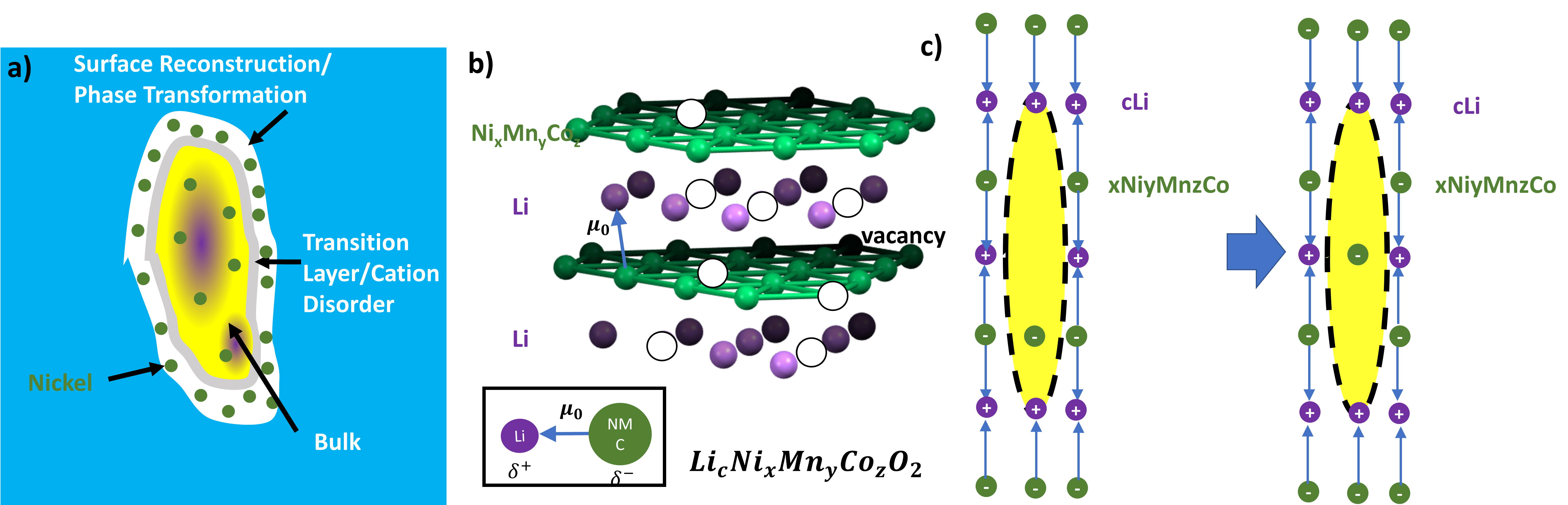}
\centering
\caption{a) Schematic of surface reconstruction and phase transformations in a cathode particle, happening from the edge of the particle to the bulk. b) Microscopic schematic of cation disorder defined by the lithium and electrons at the transition metal sites, where a nickel migrates from a transition metal layer to an empty site in a lithium layer. The dipoles are defined as from the transition metal sites to the lithium sites. c) Schematic representation of disorder in our model from layer to layer. The yellow center are the ``core'' interactions while everything outside counts as the ``bulk'' interactions.}
\label{fig:fig1}
\end{figure*}

The theory behind our model was inspired by Kosterlitz and Thouless's groundbreaking work on 2D-topological defects \cite{kosterlitz1973ordering, kardar2007statistical}. 
The defects in this model are described by ``twisting'' of ordered structures to form ``vortices''.
The idea behind our lattice model, shown in Fig. \ref{fig:fig1}a, is that for NMC defects the antisite defect is the most energetically favorable \cite{lee2012atomistic}, which makes them the driving force for phase transitions in nickel rich materials.
They can also be thought of as ``flipping'' structures that are normally topologically perfect, but through entropy and electrostatic changes from the configurational ``flip'' can have modified energies.
Since the lattice changes in these materials are quite small between the fully lithiated and empty states \cite{min2016comparative}, we do not account for lattice size changes in this material.

In Kosterlitz-Thouless (KT) transitions, the separation of the interaction energies into entropy, core interactions, and mean field interactions with the bulk is another key to the correct calculations of topological defect theory.
This is analogous to ionic Born solvation modeling, but instead of ion-water systems, our current system is a solid state system where electrostatics dominate  \cite{bashford2000generalized}.
The energy required to create a cavity in the solution is analogous to the ``core'' interaction energy, while the integrated electrostatic interactions with the bulk is analogous to the ``bulk'' of the electrostatic energies.

The energetic barrier of transitions from a perfect NMC lattice to the disordered phase using statistical models of layered lattices was captured based on the fact that electrostatic interactions dominate in these ionic crystals \cite{zheng2008effects} and the radii of nickel and lithium ions are similar in size.
The phases formed in the densified states are still uncertain from experimental measurements, but we know the phase transformations are driven by cation disorder, so the amount of defects formed can be used as an indicator for the amount of surface reconstruction or phase transformations \cite{reed2004role}.
Highly disordered nickel in the lithium layer \cite{xiao2019understanding} indicates the occurrence of spinel or rock salt phase transitions.

In the traditional Butler-Volmer (BV) phenomenological model of intercalation kinetics in battery materials, classical ion transfer is assumed to be the dominant reaction mechanism and the electronic degrees of freedom are not considered \cite{doyle1993modeling, newman2012electrochemical}.
In contrast, electron transfer is generally described using Marcus theory \cite{marcus1956theory, marcus1964chemical} for localized (metallic) electron transfer, or Marcus-Hush-Chidsey theory for delocalized electron transfer  \cite{hush1961adiabatic, chidsey1991free}, in case of Faradaic reactions at liquid/solid electrode interfaces. 
Bai and Bazant first hypothesized that electron transfer to reduce the electrode host could be rate-limiting in Li-ion batteries, compared to the fast step of lithium ion insertion, and showed the MHC theory could predict curved Tafel plots for lithium iron phosphate electrodes \cite{bai2014charge}. Fraggedakis et al. then derived a general theory of coupled-ion electron transfer (CIET) applied to lithium intercalation reaction kinetics \cite{fraggedakis2021theory} with strong experimental support from nanoscale x-ray imaging of lithium concentration evolution \cite{Zhao2022_LFP_inversion_preprint} and from pulsed electrochemical measurements for a wide range of Li-ion battery materials \cite{Zhang2022_CIET_preprint}.  In stark contrast to BV models, the CIET theory connects reaction rates to microscopic material properties and predicts curved Tafel plots with concentration-dependent limiting currents at high overpotentials. 

The basic idea behind CIET is that lithium ion transfer into the lattice is accompanied by the formation of a neutral polaron quasi-particle by electron transfer to a weakly coupled reduced state of the solid host, typically involving a reduced transition metal cation.
We theorize that ion-electron pairs are also key features of defects in intercalation materials, which may be modeled as dipoles in the dominant electrostatic free energy of the electrode. 
These dipoles can also be seen as ``bound'' defects \cite{mehrer2007diffusion}, which can be important in intermetallics.
The dipolar behavior of electrostatics \cite{jackson1999classical} interacting with the change of charges at the core are shown in this study to qualitatively reproduce the large difference in amount of defects driven at different cutoff voltages \cite{chang2000synthesis, manthiram2016nickel, manthiram2014perspective}. 

Lattice models, commonly used in physics, can be used to study layered materials, where the energy calculations are further simplified by the layered effect. 
Such models of nickel-rich  cathodes have been used to study the effect of cation disorder on voltage profiles \cite{mercer2017influence, abdellahi2016understanding}.
Nearest-neighbor lattice models have also been used to model the temperature dependent order-disorder transition of different lithium layered oxide materials \cite{gao1996changes, li1992crystal, li1994lattice}. 
However, in these models often only the nearest-neighbor or next nearest-model terms were used to study these effects, neglecting the electrostatic environment and ignoring convergence of the electrostatic interactions.
These models were also not expanded to formulate chemical potential models for dynamic free energy models.

Here, we present a general microscopic theory of cation disorder in ionic crystals, specifically applied to nickel-rich oxide degradation. 
Using dipole-charge interactions assuming a mean field made up of alternating dipole layers from the ion-electron pairs placed at the Li-TM sites as shown in Fig. \ref{fig:fig1}b, we verify the convergence of these electrostatic calculations (shown in Appendix B) and present a rigorous electrostatic mean field model.
These dipole interactions have been theorized to play a role in the formation of these disordered materials \cite{garcia2020strain, clement2020cation, knauth2002solid}.
Then, accounting for configurational entropy, a free energy model for the material is derived, which can be used to derive chemical potentials for dynamic simulations.
This model is derived mainly for disorder and not for intercalation, but a verification of the chemical potential reveals that the voltage range is on the same order as a battery intercalation material, as shown in Fig \ref{fig:fig3}c, further validating our results.

We formulate this model as a first order approximation to be used as a theoretical model for any layered oxide cathode material disorder that one wishes to study. 
The model can also be extended to materials such as lithium iron phosphate \cite{malik2010particle} where analogous iron-antisite defects have found to be important in determining the particle-size-dependent effective diffusivity.
The ideas based in electrostatics and statistical thermodynamics come together to formulate a model for studying degradation of lithium oxide materials.

\section{Theory}\label{sec:body}

\subsection{Bulk}

\subsubsection{Chemical Potential}

The model is formulated as follows.
Based on the alternating locations of the nickel and lithium layers, dipoles of alternating directions are formed between the nickel and lithium layers in the mean field approximation, where ``even'' and ``odd'' layers have dipoles in alternating directions, as shown in Fig. \ref{fig:fig1}c.
However, in addition to the fact that the bulk dipole layers alternate directions when a nickel atom migrates to a lithium layer, the core part of the interaction is also modified when a defective configuration is formed.
We account for this similarly to Kosterlitz and Thouless \cite{kosterlitz1973ordering}, where these energy calculations are separated into a ``core'' term and a ``bulk'' term, where the bulk term is accounted for from the alternating layers and the core term is from the charge interactions near the defective site.

To calculate the difference between energies of the defective and normal configurations \change{$\Delta G = G_1 - G_2 = H - T S$}, we find the difference between the two states--a defective configuration where a nickel atom has moved to a lithium site \change{$G_2$}, and a non-defective configuration where the nickel atom is in its original nickel site \change{$G_1$}, based on the crystal structure of LiNiO$_2$ from the Cambridge Structural Database \cite{groom2016cambridge, macrae2020mercury}.
The mean field dipoles can be defined at different concentrations of lithium $c$, defect concentrations of nickel $v$, and ratios of nickel manganese cobalt $x:y:z$.
\change{The variable names used in this paper are redefined in Appendix \ref{appdx:symbols} for clarity.}
The dipoles in the structure are shown in Fig. \ref{fig:fig1}b using electronegativity, defined from the positively charged lithium sites (from the lithium ion placed in the sites) to the negatively charged nickel-manganese-cobalt sites (from the electron localized on the transition metal), with the dipole written as $\change{\boldsymbol{\mu}_{0}} = e\change{\mathbf{r}_{0}}\left((x-2v)\text{EN}_{Ni}+y\text{EN}_{Mn}+z\text{EN}_{Co}-\text{EN}_{Li}\right)$, where \change{$\mathbf{r}_{0}$} is the vector distance between transition metal site and lithium site, and EN is the electronegativity of the atom or the ``attractiveness'' of an atom to electrons \cite{israelachvili2015intermolecular}.
\change{We see that there are alternating rows of dipoles pointing in opposite directions from the formulation of dipoles in this model.}
Only the dipoles added by intercalation are considered in this crystal structure and not the ions pre-existing in the non-intercalated structure, since when an energy difference is calculated between the original and defective state, the intercalation host crystal interactions cancel out.
To avoid the fact that many electronegativity scales are based on formation energies, the Allred-Rochow scale was chosen.
This scale is more simplistic than other electronegativity scales and generally uses the ideas of electrostatics based on effective nuclear charge, consistent with our theory \cite{allred1958scale}.

In the lithium layers and transition metal layers, there are automatic constraints on the concentrations of lithium and defective nickel in the lithium layers, or nickel and vacancies in the transition metal layers.
The definition of the material ratio gives us that $x + y + z = 1$, and from the site constraint of lithium atoms, we know that $c+ v \leq 1$.
By the mass constraint of nickel atoms, $v \leq x$ must always be true.
These constraints are automatically satisfied by how the entropy equation was defined.

We consider the perfect configuration \change{$G_1$} before a defect is formed in the core as well as a defective configuration \change{$G_2$}. 
Assuming that the neutral lattice contains a lithium-electron pair for notation purposes, the reaction that occurs can be written in Kroger-Vink notation as
$\mathrm{V_{Li}^{'}} + \mathrm{Ni_{Ni}^x} \rightarrow \mathrm{Ni_{Li}^{\cdot}} + \mathrm{V_{Ni}^{''}}$.
For comparison, the lithium intercalation reaction written in Kroger-Vink notation is
$\mathrm{e^{'}} + \mathrm{V_{Li}^{'}} + \mathrm{Li^{\cdot}} + \mathrm{Ni_{Ni}^{x}} \rightarrow \mathrm{Li_{Li}^{x}} + \mathrm{Ni_{Ni}^{x}}$.
Since the oxygen lattice around lithium and nickel ions are identical and the radii of lithium and nickel are very similar, by symmetry the oxygen atoms can be neglected in the modeling of the crystal structure and only the lithium and nickel layers are considered.
The crystal structure used was taken from the the Materials Project structure mp-632864 for LiNiO$_2$ \cite{jain2013a, jain2013b, laubach2009, hirano1995, adipranoto2014}.

In this structure, dipoles are formed from the intercalated lithium ion, which is positively charged, interacting with the localized electron on the transition metal ion.
Thus, we have alternating layers of dipoles that form the bulk of the electrostatic free energies of the crystal structure.
When an antisite defect is formed, one of the dipoles is broken and forms a broken ``core,'' \change{which is also called the defect core for future reference}.
When studying the energetic interactions of antisite defect formation, we can separate the interactions into the broken ``core'' term as well as the electrostatic bulk interaction terms, shown in Fig. \ref{fig:fig1}c where the circled yellow site is the broken core and the outside is the mean field term.

We first start by studying the bulk interaction term, also known as the mean field (MF) term.
It simplifies the problem to consider splitting the dipole layers between the ones pointing ``up'' and the ones pointing ``down'' by the symmetry between these layers.
In the following notation, ``even'' indicates the layer that a core electron would belong to if it was moved half a layer up and the alternating layers pointed downward in the diagram, while ``odd'' indicates a layer that the core electron would belong to if it was moved half a layer down and the alternating layers pointed upward in the diagram.
The definition of the dipole is thus always positive in ``odd'' layers and negative in ``even'' layers in the original configuration, while it is always negative in ``odd'' layers and positive in ``even'' layers in the defective configuration as shown in Fig. \ref{fig:fig1}c \cite{niklasson2006electronic}.
The electrostatic interactions in the layers are thus opposite to each other.
By the definition of charge-dipole interactions, the even and odd interactions will be 
\begin{equation}
\begin{split}
    H_{MF,even} & = -\frac{1}{2}\frac{cqe}{4\pi \varepsilon} \sum_{i \in even} \sum_{j \in i} \change{\frac{-\boldsymbol{\mu}_{0}\cdot \boldsymbol{r}_{ij}}{r_{ij}^3} }\\
    H_{MF,odd} & = -\frac{1}{2}\frac{cqe}{4\pi \varepsilon} \sum_{i \in odd} \sum_{j \in i} \change{\frac{\boldsymbol{\mu}_{0}\cdot \boldsymbol{r}_{ij}}{r_{ij}^3}}, \\
\end{split}
\end{equation}
where the factor of $\frac{1}{2}$ accounts for the fact that the charges are split over a dipole in the layer above and the layer below, and the $-cqe$ prefactor accounts for the amount of dipoles at each site in a mean field description, since only the intercalated lithium sites $c$ can have lithium-electron dipoles \cite{bottcher1973theory, jackson1999classical, landau2013electrodynamics, yamamoto2008charge}.
Here, $e$ is a unit charge and $q$ is the magnitude of the charge.
We sum over the layers in the crystal structure indexed by $i$, which are separated into even and odd layers, and then sum over the sites in the $i$th layer indexed by $j$.
Since the dipole vector, \change{$\boldsymbol{\mu}_{0}$}, is defined as pointing from the transition metal layer to the lithium layer, assuming the centered atom is the red lithium site, the even layers are shown in the image are the orange dipole layers, while the odd layers are the blue dipole layers. 
Here, the distance \change{$\boldsymbol{r}_{ij}$} is the distance between the center of the dipole \change{$\boldsymbol{\mu}_{0}$} and the \change{defect center}, and the scalar \change{$r_{ij}$} is the magnitude of that vector.

Because ions in this problem are assumed to only move as a result of defects or intercalation and the induced dipoles from electronic movement are ignored, the dielectric constant applied is the static dielectric constant of this material.
These approximations cause our simple theoretical model to neglect induced many-body interactions that are not captured by a mean-field model, but it is a good first approximation.
The full mean-field theory is then written as
\begin{equation}
\begin{split}
    H_{1,MF} & = \left(H_{MV_{even}} +H_{MV_{odd}}\right) \\
    H_{2,MF} & = - \left(H_{MV_{even}} +H_{MV_{odd}}\right),
\end{split}
\end{equation}
split over the even and odd layers.

\begin{figure}[t]
\includegraphics[width=0.5\textwidth]{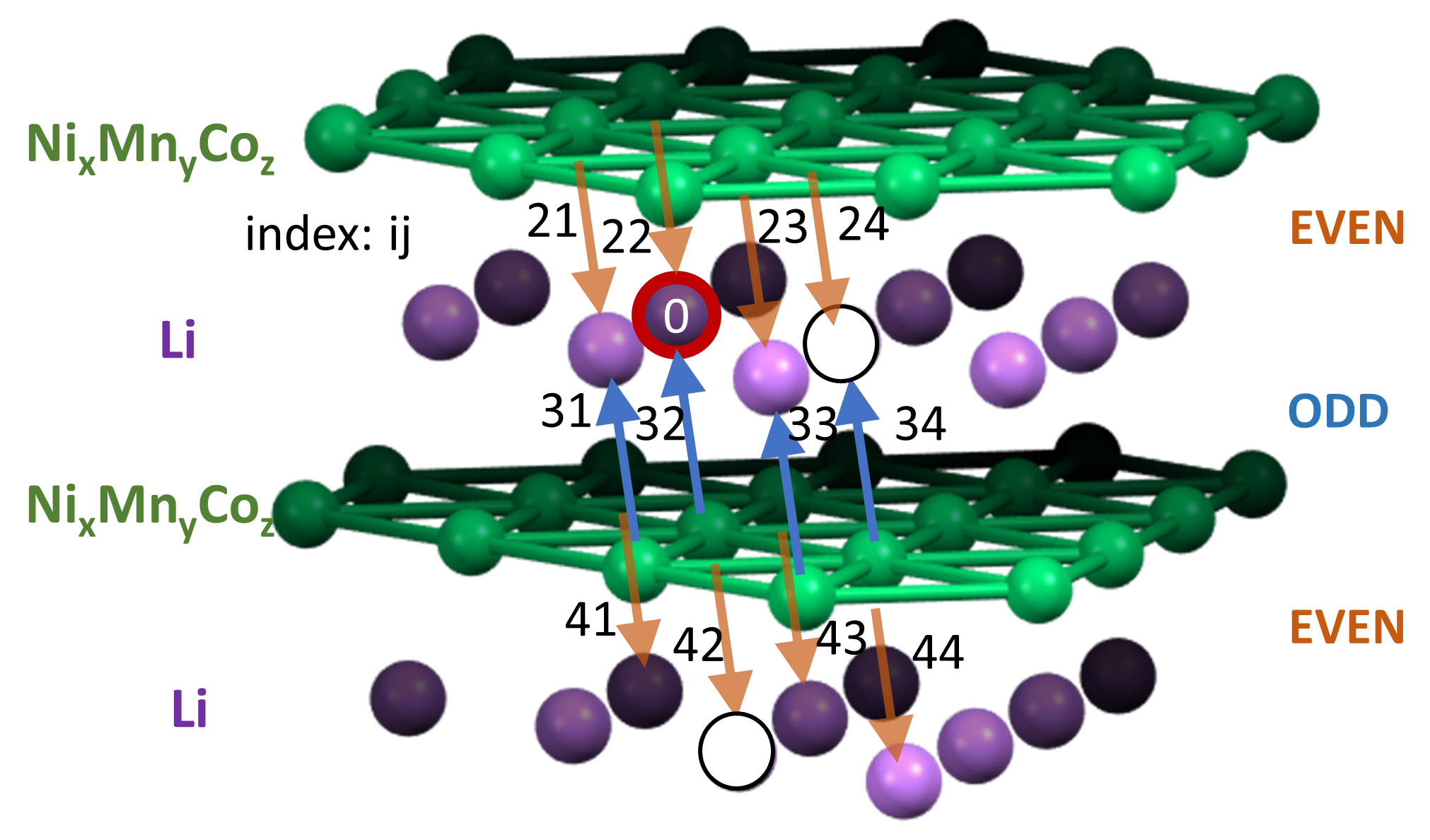}
\centering
\caption{Detailed schematic of charge-dipole model, where in $\boldsymbol{\mu}_{ij}$, $i$ is the layer index, which can be either ``even'' or ``odd'', and $j$ is the atom number in the layer. ``0'' is the central site where the antisite defect occurs.}
\label{fig:appendix2}
\end{figure}

The broken ``core'' of the structure is studied next.
It is known that the defective nickel is more likely to be in the reduced (+3) oxidation state.
Since at the core, there is only one negatively charged transition metal interacting with the mean field, this greatly simplifies our energy calculation to be interactions with the two lithium ions within the vicinity of the layer, shown in Fig. \ref{fig:fig1}c.
We use charge-charge interactions to model the core interactions.
In the core, only half of the ions on the edges are considered (the other half is used to generate the dipoles in the mean field terms).
Thus, at each end of the core, we have a nickel-lithium interaction from the nickel to the edge.
The electrostatic core interactions in the two configurations are found to be
\begin{equation}
\begin{split}
    H_{1,core} & = -\frac{(1-v)}{2} \left( \frac{c(qe)^2}{4\pi \varepsilon r_0} + \frac{c(qe)^2}{4\pi \varepsilon (3r_0)}\right)\\
    H_{2,core} & = -\frac{(1-v)}{2} \left( \frac{c(qe)^2}{4\pi \varepsilon (2r_0)} + \frac{c(qe)^2}{4\pi \varepsilon (2r_0)}\right).\\
\end{split}
\end{equation}
This gives us
\begin{equation}
\begin{split}
     H_{MF} & = H_{2,MF} - H_{1,MF}\\
    H_{core} & = H_{2, core} - H_{1, core}
\end{split}
\end{equation}
as the final enthalpic interaction \change{energy difference between the defective and bulk configurations}.

\begin{figure*}[t]
\includegraphics[width=\textwidth]{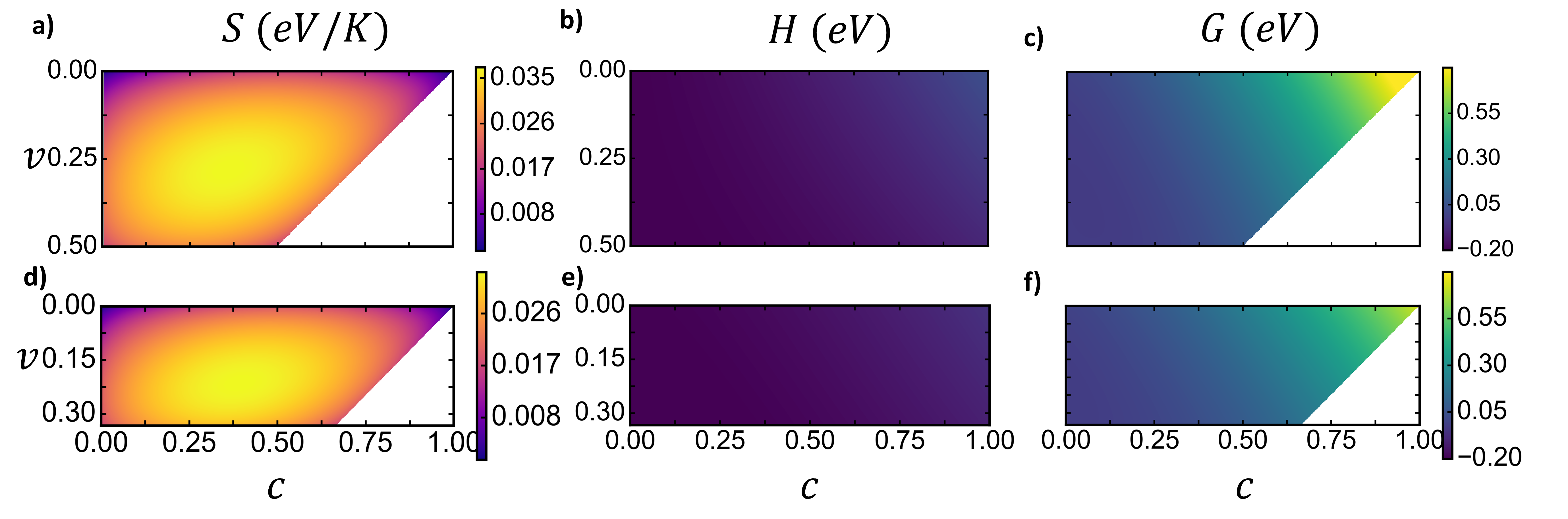}
\centering
\caption{a), b), and c) are the entropy, enthalpy, and Gibbs free energy calculations for NMC532, while d), e), and f) are the entropy, enthalpy and Gibbs free energy calculations for NMC111. Because of the total site constraint in the lithium layer in the crystal structure, when the concentration in the lithium layer is above a certain level, there are not enough sites for nickel to have a high concentration of defects in the system, so there are white triangular regions of no solution for the entropy calculations. This also causes there to be regions of no solutions for the free energy calculations.}
\label{fig:fig2}
\end{figure*}

By using the definition of dipole charge-interactions \cite{jackson1999classical}, the final mean field energy difference between the two configurations is found to be 
\begin{equation}
    H_{MF} = - \frac{cqe}{4\pi \varepsilon} \left( \sum_{i \in \text{even}} \sum_{j \in i}\change{ \frac{\boldsymbol{\mu}_{0}\cdot \mathbf{r}_{ij}}{|r_{ij}|^3}} -  \sum_{i \in \text{odd}} \sum_{j \in i} \change{\frac{\boldsymbol{\mu}_{0}\cdot \mathbf{r}_{ij}}{|r_{ij}|^3}}\right),
\end{equation}
is the vector between the center of dipole $ij$ and the \change{defect center} and \change{$\boldsymbol{\mu}_{0}$} is the dipole between lithium sites to transition metal sites defined in the previous solution, where \change{$r_{ij} = |\mathbf{r}_{ij}|$}.
The defect core interaction difference is found to be 
\begin{equation}
   H_{core} = \frac{(qe)^2 (1-v)c}{12\pi \varepsilon r_0},
\end{equation}
The final electrostatic energy is found to be 
\begin{equation}
\label{eq:enthalpy}
    H = H_{MF}+ H_{core},
\end{equation}
shown in Figs. \ref{fig:fig2}b and \ref{fig:fig2}e.
The electrostatic energies in these materials were found to be lower at a state with more defects, indicating that a highly defective state is energetically favorable.
This is expected as the formation of such a state reduces the magnitude of electrostatic interactions between layers.

In this electrostatics problem, the dielectric constant $\varepsilon$ is estimated using the Clausius-Mosotti relation for a spherical inclusion in a homogeneous effective medium.
Since the dielectric response of the material is based on the interactions induced by other atoms and the energies converge quickly in these structures (shown in Appendix B), in the core and near field interactions, a simple dielectric constant can be used in this model.
The movement of the ionic lattice is considered in this problem, so the low frequency (static) dielectric constant needs to be applied.
The dielectric constant can be found to change with the amount of intercalation or defects locally.
When the crystal structure is nearly perfect, the dielectric constant for layered lithium oxide materials has been found to linearly decrease with the increase of defect concentration, specifically for lithium niobium oxide \cite{iyi1992ceramics, palatnikov2000effects, xue2002dielectric}.

We seek an approximation rule to calculate the average dielectric constant of materials at different concentrations and to reproduce the behavior seen for lithium metal oxide materials.
The dielectric constant of metal oxides is estimated following the additive rule in ``well-behaved'' metal oxides, based on Clausius-Mosotti dielectric theory \cite{shannon1993dielectric, coker1976empirical}, which has been found to work extremely well for many kinds of oxide materials \cite{wilson1965dielectric, jonker1947dielectrische}.
The additive rule has been found to be a good predictor of the dielectric constant of oxide materials even before measurement, with the dielectric constant $\varepsilon$ obtained from
\begin{equation}
    \frac{\varepsilon-1}{\varepsilon+2} = \frac{4}{3}\pi\sum_{i}\alpha_i n_i,
\end{equation}
where $\alpha_i$ is the polarizability of the atom $i$ and $n_i$ is the number density of atom $i$.
For our material, we consider atom types Li, Ni, Mn, Co, and O.
The dielectric constant decreases with respect to lithium concentration and defect concentration, as shown in Appendix C.
This dielectric constant calculation is only valid for bulk dielectric constants since we do not consider image charge effects from the bulk/electrolyte interface.

The second part of the energy from the configurational entropy of the model can be split into two parts.
First, in the lithium layers, the sites are either filled with lithium, defective nickel, or empty, which can be written as
\begin{equation}
    \Omega_{Li} = \frac{N!}{(Nc)!(Nv)!(N(1-c-v))!},
\end{equation}
where $N$ is the total number of lithium sites considered.
In addition, the nickel layers are either filled with nickel or empty, with the following number of combinations
\begin{equation}
    \Omega_{Ni} = \frac{(Nx)!}{(Nv)!(N(x-v))!}.
\end{equation}
The total configurational entropy change is described as 
\begin{equation}
\label{eq:entropy}
\begin{split}
S = &  k_B \ln{\left( \Omega_{Li}*\Omega_{Ni}\right)} \\
\approx & k_B N\left( x\ln{x} - c\ln{c} - (1-c-v)\ln{(1-c-v)} \right.\\
& \left. -2v\ln{v}-(x-v)\ln{(x-v)}\right),
\end{split}
\end{equation}
which is plotted in Figs. \ref{fig:fig2}a and \ref{fig:fig2}d for NMC532 and NMC111 separately.
The configurational free energy prefers moderate values for $v$ and $c$ because of the higher number of possible states at at these concentrations.
More importantly, the entropy of the configuration limits the accessible states at higher lithium concentrations, because it is physically impossibly for high concentrations of defects to be reached at high lithium concentration from lack of available sites, creating the inaccessible triangular regions in Fig. \ref{fig:fig2}a and d.

\begin{figure*}[t]
\includegraphics[width=5in]{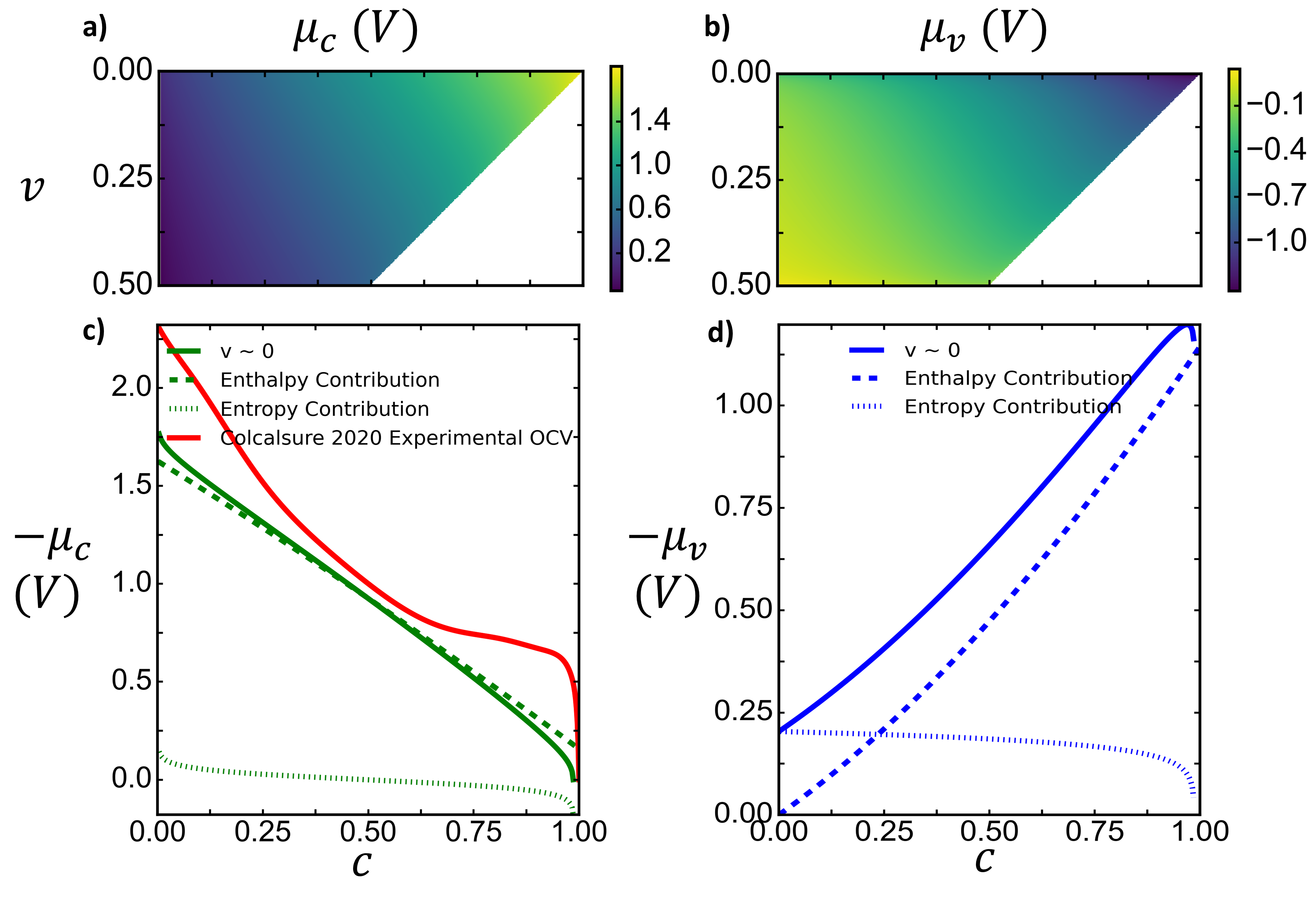}
\centering
\caption{a) Intercalation chemical potential for a NMC532 material. b) Defect chemical potential for a NMC532 material. c) Intercalation chemical potential when defect amount is close to zero and the contributions from enthalpy and entropy. We also plot an open circuit voltage from experimental measurements \cite{colclasure2020electrode} for comparison. d) Defect chemical potential when the defect concentration is close to 0 and the contributions from entropy and enthalpy. (Potentials are shifted in c) and d) by an arbitrary reference potential for ease of readability.)}
\label{fig:fig3}
\end{figure*}

The final free energy can be described as $ G =  H - T  S$ from Eqs. \ref{eq:enthalpy} and \ref{eq:entropy}, plotted in Figs. \ref{fig:fig2}d and \ref{fig:fig2}f.
\change{More information on these calculations can be seen in Appendix \ref{appdx:numerical}.}
This free energy is dominated by the electrostatics, but there is a strict cutoff from the possible available sites in the entropic component.
The diffusional chemical potentials, which describe the dynamic behavior of the model, can be defined from the free energy.
The diffusional chemical potentials are defined as $\mu_c = \frac{\delta G}{\delta c}$ for intercalation and $\mu_v = \frac{\delta G}{\delta v}$ for defects, shown in Fig. \ref{fig:fig3}, and the analytical solutions for the chemical potentials can be found in Appendix A.
From Figs. \ref{fig:fig3}b and \ref{fig:fig3}d, at lower concentration, there is a larger driving force towards a more defective state, qualitatively matching experimental measurements where the high voltage/low concentration regions cause more cation disorder.
Since the chemical potential is defined with an arbitrary reference value, for ease of comparison between the intercalation and the defect chemical potential we choose to shift the reference potentials to overlap.
The chemical potential contribution from entropy is derived analytically in the appendix, while the contribution from enthalpic terms is calculated numerically in Appendix A, and the contributions are plotted in Figs. \ref{fig:fig3}c and \ref{fig:fig3}d.
For both intercalation and defect formation, the enthalpic chemical potential dominates the trend in these materials.

\begin{figure}[t]
\includegraphics[width=0.45\textwidth]{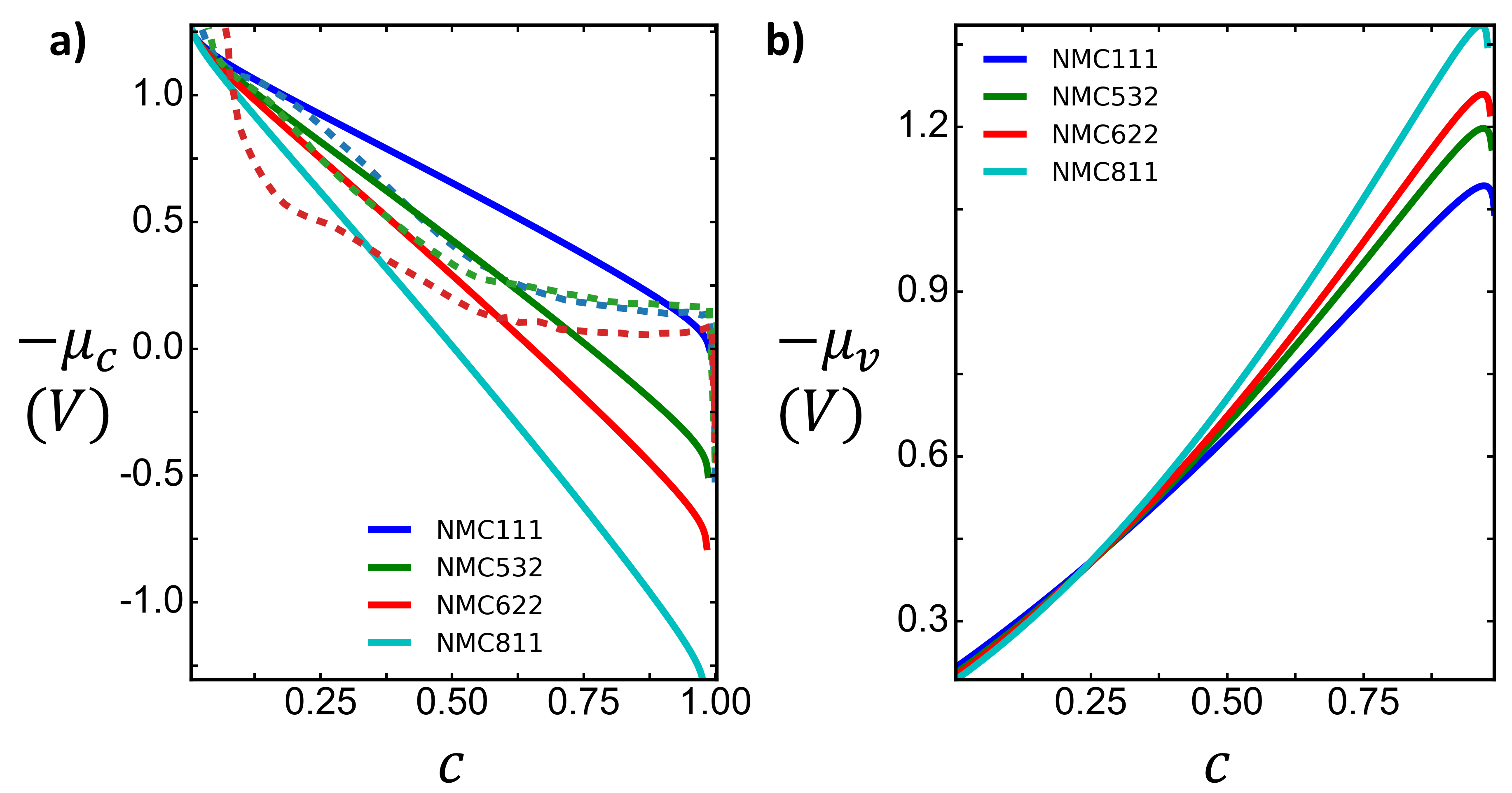}
\centering
\caption{Nickel ratio effect on the a) intercalation and b) defect formation chemical potentials of the material. The solid lines are for the calculated values, while the dotted lines are from experimentally measured values from Ref. \cite{jung2017oxygen}.}
\label{fig:fig6}
\end{figure}

The advantage of these simplistic models is that they can be applied to nickel rich materials of any composition and provide a rough estimate of the energy of any of these new compositions of materials in Fig. \ref{fig:fig6}.
In the previous discussion, Figs. \ref{fig:fig2}c and \ref{fig:fig2}f provide an example of different ratios of nickel in the crystal structures for NMC532 and NMC111.
Experimentally, higher nickel ratio materials were revealed to have larger voltage ranges \cite{jung2017oxygen} and degrade more quickly than lower nickel ratio materials \cite{noh2013comparison}, which is also shown in our theoretical predictions for chemical potential of these materials in Fig. \ref{fig:fig6}.
Our model follows the general trend of increasing nickel content causing a larger slope in the chemical potential for intercalation. 
The different electronegativity of the transition metal materials, which modifies the dipole magnitude, causes this change in slope.
In addition, the nickel concentration plays an important role in the entropy cutoff by controlling the maximum amount of nickel in the material that can form defects.
Though the enthalpic interactions dominate, the entropic interactions also become important in restricting the strict limits of amount of defects.
At lower concentrations, the driving force towards defect formation is higher for larger nickel content materials, which has been observed experimentally as well \cite{jung2017oxygen}.
We note that though the model for ``intercalation'' description was not as complex, we still capture the effect of the transition-metal concentration on the intercalation chemical potential of the material.
The slope changes near delithiated and fully lithiated for lithium intercalation are expected to come from different ionization potentials of the lattice since the lithium-rich or poor regions exist in a highly charged state \cite{sushko2013oxygen}.

\subsubsection{Dynamic Model} \label{sec:rxn_diff}

Using the free energy models formulated above, chemical potentials, which are the driving forces for dynamic models, can be derived for the material.
We model the most simplistic version of a battery model--a single particle model.
The dynamic model at the surface is described using a single particle reaction-diffusion model with a driving force from the diffusional chemical potential gradient.
Both the lithium concentration $c$ in the lithium layer and the defect concentration $v$ in the lithium layer are modeled through nonequilibrium thermodynamic driving forces, the gradients of chemical potential \cite{kondepudi2014modern, de2013non}.
The boundary conditions applied for both models are applied through the intercalation reaction for lithium concentration and the oxygen degradation reaction at the surface, described in the following section.

\change{A simple single particle model \cite{guo2010single} was used in this system to model the cycling behavior, based on more complicated porous electrode models \cite{newman2012electrochemical}.}
For intercalation, the simple form of mass conservation with a diffusive driving force can be described as 
\begin{equation}
\label{eq:cons_c}
    \frac{\partial c}{\partial t} = - \nabla \cdot \mathbf{F}_c,~ \mathbf{F}_c = - \frac{D_cc}{k_BT} \nabla \mu_c,
\end{equation}
where $D_c \propto (1-v)$ because of the effect of ``blocked'' sites on the diffusivity of the material \cite{malik2010particle, kang2006factors}.
In addition, because of the change of the maximum number of required sites, during cycling, the chemical potential parameter actually varies with $\mu_c(c/(1-v))$ to rescale to the proper number of total sites.
The dynamic equation for conservation of defects is described as 
\begin{equation}
\label{eq:cons_v}
    \frac{\partial v}{\partial t} = - \nabla \cdot \mathbf{F}_v,~ \mathbf{F}_v = - \frac{D_vv}{k_BT} \nabla \mu_v,
\end{equation}
if we assume a constant diffusivity coefficient for both models for the flux $\mathbf{F_{c/v}}$ in the bulk.
For simplicity, we assume that the diffusion coefficients are constant.
The boundary equations applied for the particle are
\begin{equation}
\label{eq:BC_c}
    -\mathbf{n} \cdot \mathbf{F}_v = i_{int}
\end{equation}
for surface reaction of intercalation $i_{int}$. \cite{bazant2017thermodynamic, bazant2013theory}
For the defect concentration, a simple degradation mechanism explained in 2.2.2 is prescribed, which gives the boundary condition of 
\begin{equation}
\label{eq:BC_v}
    -\mathbf{n} \cdot \mathbf{F}_c = 2i_{oxy}.
\end{equation}
\subsection{Surface}

\subsubsection{Intercalation}

Since blocked sites from defects play a central role in our theory, it is imperative to use an accurate reaction model that captures configurational entropy and polaron formation energies. 
For this purpose, the recent theory of coupled ion-electron transfer (CIET) for ion intercalation \cite{fraggedakis2021theory} is adopted as the boundary condition, where the concept of ion-electron polaron pairs complements the idea of dipole pairs to describe the electrostatic interactions among disordered cations.
In CIET theory, the blocked sites play a much more dominant role than in classical Butler-Volmer models, because theory predicts a reaction-limited current $i_{lim}$ which has a strong asymmetry dependence on all of the species concentrations: $i_{lim} = c_+(1-c-v) i_r^*$ for negative overpotentials and $i_{lim} = c(1-c-v)i_r^*$ for positive overpotentials \cite{Zhang2022_CIET_preprint}. 

These limits arise from the general form of the intercalation rate given by 

\begin{equation}
   i_{int} =  i_r^* \int_{-\infty}^\infty (1-c-v) \left(c_{+}n_e p_{red}(\varepsilon) - c(1-n_e) p_{ox}(\varepsilon)\right)\rho  d\varepsilon ,
\end{equation}

where the conditional probability that an electron of energy $\varepsilon$ relative to the Fermi level participates in reduction or oxidation is given by 
\begin{equation}
    p_{red/ox}(\varepsilon) = \frac{1}{\sqrt{4\pi\lambda/k_B T}} \exp{\left(- \frac{(\lambda \pm \eta_f \mp \varepsilon)^2}{4\lambda k_B T} \right)}.
\end{equation}
Here, $c$ is the dimensionless lithium ion concentration (filling fraction) in the host crystal; $v$ is the dimensionless defect concentration, or the nickel filling fraction in the lithium layers; $c_+$ is the lithium ion concentration at the reacting surface, related to the nearby electrolyte concentration by an adsorption isotherm, assuming fast surface adsorption compared to CIET intercalation; $n_e(\varepsilon)$ is the Fermi-Dirac distribution, and $\eta_f$ is the formal overpotential defined as $e\eta_f = e\eta + k_B T\ln{\frac{c_+}{c}}$ \cite{Zhang2022_CIET_preprint}.
\change{As the overpotential is $e\eta = \mu_R - (\mu_O + n \mu_e)$ \cite{bazant2013theory} where $n$ is the number of electrons}, we see that the overpotential $e\eta = \mu_c - k_B T \ln{c_+} + e\Delta \phi$, is related to the difference between the intercalation chemical potential $\mu_c$ and the potential difference $\Delta \phi = (\phi_e-\phi)$ between the solid $\phi_e$ and the electrolyte $\phi$.
The parameters in the model are $\lambda$, the Marcus reorganization energy for electron transfer; $i_r^*$, the prefactor for current related to electronic coupling and the ion-transfer energy; $\rho(\varepsilon)$, the energetic density of states (band structure).

Our reaction rate differs from the typical CIET model for lithium intercalation since the empty sites must also be reduced by the number of blocked sites in the material from defects, contributing to the factor by $v$.
Under the assumption that the electron donor is metallic, we can assume a uniform density of state \cite{fraggedakis2020tuning, chidsey1991free} which allows us to use the simple and accurate approximation of the MHC formula by Zeng et al. \cite{zeng2014simple} to derive a closed formula for the CIET reaction rate for lithium insertion:
\begin{widetext}
\begin{equation}
    i_{int} = i_r^* (1-c-v) 
    \left(\frac{c_{+}}{1 + \exp{\left(\frac{\eta_f}{k_BT}\right)}} - \frac{c}{1+\exp{\left(\frac{-\eta_f}{k_B T}\right)}} \right)\text{erfc}\left(\frac{\lambda - \sqrt{1+\sqrt{\lambda} + \eta_f^2}}{2\sqrt{\lambda k_B T}} \right),
\end{equation}
\end{widetext}
which reduces to the form given by Zhang et al. \cite{Zhang2022_CIET_preprint} in the limit of a defect-free host crystal, $v=0$.

\subsubsection{Defects}

\begin{figure*}[t]
\begin{center}
    \includegraphics[width=6.5in]{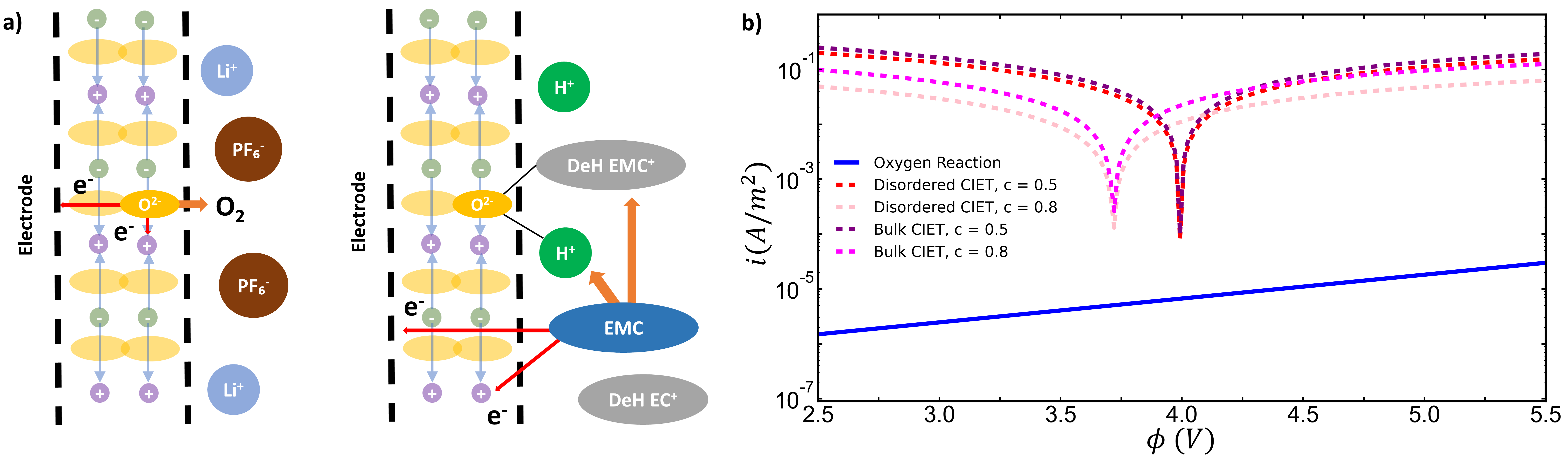}
\end{center}
\caption{a) Schematic of oxygen reaction and change of dielectric constant at the surface of the particle, where the yellow circles are oxygen ions. Nickel is driven into the bulk crystal, and the oxygen reaction at the electrolyte interface forms oxygen vacant sites, free electrons, and releases oxygen gas or other oxidation products to the electrolyte. \change{The mechanisms of oxygen release and dehydrogenation of EMC/EC are both proposed in this model \cite{jung2017oxygen, zhang2020revealing}.} b) Evans diagram of the oxygen reaction model used and behavior relative to voltage applied. The coupled-ion electron transfer intercalation reactions are plotted as well at bulk ($v=0$) and defective ($v=0.1$) phases with different lithium concentrations. It can be seen that defective phases reduce the magnitude of the intercalation reaction through blockage of available sites. 
}
\label{fig:fig4}
\end{figure*}

In the bulk, phase transformations and cation disorder are triggered by the oxidation of reactive oxygen ions at the solid surface, \change{ typically an edge plane of the layered oxide crystal. } \cite{yan2019injection, zheng2017suppressed}.
At high voltages, degradation at the surface is much more pronounced, especially those triggered by oxygen vacancy formation and oxygen changes at the surface \cite{xu2011identifying, breger2006effect, mohanty2015understanding}; experimentally, oxygen vacancy formation at high voltages has also been observed \cite{jung2018temperature}.
Since the dielectric ``bulk'' of the medium consists of oxygen ions, and the lithium and nickel ions only interact at close range for quick convergence of the free energies, we assume that the bulk dielectric constant is affected by the local oxygen, defect, and lithium concentrations, where the oxygen vacancies are assumed to not propagate into the bulk, as shown in Fig. \ref{fig:fig4}a \cite{xiao2019understanding}.
It is generally considered that oxygen evolution only happens at the surface of nickel rich electrodes because of the large migration barrier of the oxygen in the bulk \cite{xu2020challenges, kong2019kinetic}.
Recent experiments, however, have suggested that oxygen vacancies can propagate into the bulk \cite{csernica2021persistent, yan2019injection}.
After the first couple cycles, bulk induced oxygen vacancy degradation can start influencing the degradation of overlithiated oxides \cite{yan2019injection, csernica2021persistent}.
For simplicity, we can assume that oxygen vacancies propagate slowly into the bulk and account for them at the surface only.
As such, we will have an oxygen boundary condition only occurring at the surface.

\change{The loss of available oxygen at the interface can be explained by several possible mechanisms, sketched in Fig. \ref{fig:fig4}a.
In the simplest possible mechanism,} oxygen ions are oxidized at the surface according to the half reaction,
$\mathrm{O^{2-}} \rightarrow 0.5 \mathrm{O_2} + 2\mathrm{e^-}$,
and there is a loss of oxygen ions at the surface in the lattice \cite{yan2019injection} that can be released as gases \cite{jung2017oxygen}, which has been experimentally observed.
\change{Another possible mechanism involves oxidation and dehydrogenation of the organic electrolyte solvent, which has been observed experimentally~\cite{zhang2020revealing, karayaylali2019coating}. We propose that this reaction could release electrons to the crystal that trigger cation disorder while creating a reactive hydroyxl group on the surface, again involving the reactive oxygen ion at the edge plane.
The reaction can be modeled as $\mathrm{EMC/EC} \rightarrow \mathrm{DeH~EMC/EC^+} + \mathrm{H^+} + 2 \mathrm{e^-}$, where the proton and dehydrogenated electrolyte product can bond with the oxygen at the surface.
The products have been experimentally observed through FT-IR experiments for EMC and EC.
}

Experimentally, we know that the amount of oxygen degradation significantly affects the amount of cation disorder \cite{armstrong2006demonstrating, qian2014uncovering}.
\change{For the oxygen release mechanism, }we postulate that when oxygen is oxidized in a material, the loss of two oxygens in a bulk requires the solid to accept electrons to conserve charge neutrality, which occurs through oxidation of transition metals.
\change{One electron is required for the electrochemical reaction, and one is required for the oxidation of the nickel ion to +3 state.}
The full reaction can be written in Kroger-Vink notation as
$2\mathrm{Ni_{Ni}^{\cdot}} + \mathrm{O_O^x} \rightarrow 2 \mathrm{Ni_{Ni}^x} + \mathrm{V_O^{\cdot \cdot}} + 0.5 \mathrm{O_2}$.
Thus, the amount of oxygen degradation increases the concentration of the reduced nickel (+3) in the material, which heightens the possibility of cation disorder.
\change{For the dehydrogenation mechanism, we propose that dehyrogenation releases two electrons, one which is required at the electrode and the other which oxidizes the nickel ion.
Thus, the amount of reduced nickel in the material also increases, causing a similar effect as the gas release mechanism to the cation disorder in the system.}

For the simplest model, we assume the dependence of the reactant, the reduced nickel, in the defect formation reaction is linear.
We know that $i_{v}$ is proportional to the amount of oxygen loss, which is $i_{oxy}$.
Thus we apply $i_v = i_{oxy}$ to the defect formation conservation  equation boundary equation to obtain \change{$-\mathbf{n} \cdot \mathbf{F}_c = i_{v} = i_{oxy}$}.
This simple boundary condition which includes the electrochemical surface reaction components of our degradation can be applied to the model, while keeping the focus on the bulk defects triggering phase transformations.
The formation voltage for this reaction is roughly $E^\theta= \SI{4.4}{V}$, which also depends on the ratios of transition metals as well as the electrolyte used \cite{zheng2008effects, armstrong2006demonstrating}.
Therefore, the overpotential driving this reaction is \change{$e\eta = E^\theta + e\Delta \phi$}.

Based on the observation that oxygen formation reaction is irreversible, a simple Tafel reaction model was used \change{for the oxidation current
\begin{equation}
    i_{oxy} = 
     k_{0,oxy} c_{oxy} \exp{\left(\frac{\eta}{k_B T}\right)},
 \end{equation}}
with a reaction parameter of exchange current density $k_{0,oxy}$, shown in Fig. \ref{fig:fig4}b, assuming that the increase of oxidation products (e.g. oxygen gas) is usually released to the environment \cite{bazant2017thermodynamic}.
$c_{oxy}$ is the concentration of oxygen ions at the solid surface, which can be modeled by the conservation equation in Appdx. \ref{appdx:numerical}.
The magnitude of reaction drops off as $\eta \leq 0$, and increases as $\eta > 0$, and the amount of reaction decreases linearly as the amount of oxygen vacancies increase.

\section{Simulations}\label{sec:simulations}

\subsection{Cycling}

\begin{figure*}[t]
\begin{center}
    \includegraphics[width=6.5in]{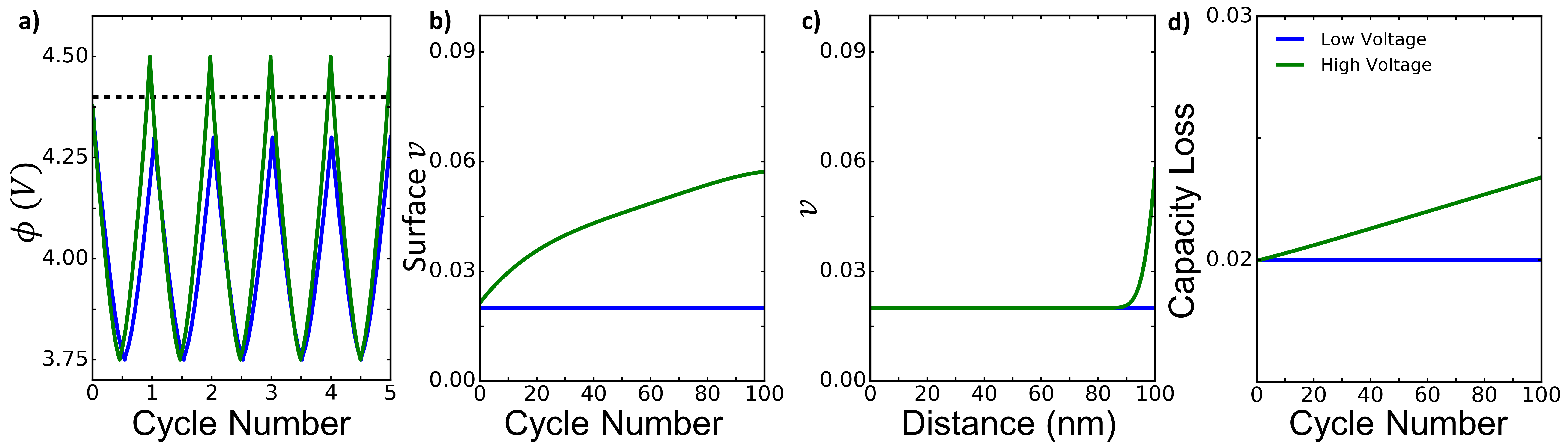}
\end{center}
\caption{a) Voltage plots for the first five cycles between low and high voltage simulations for NMC532. Higher voltage cycling tends to exceed the potential for oxygen formation. b) Defect growth in the first 100 cycles at the surface of the particle. c) Defect growth at the end of the first 100 cycles throughout the length of the particle, where $R = 0$ is the center of the particle and $R = \SI{100}{nm}$ is the edge of the particle. d) Capacity loss during cycling with voltage limits of either $\SI{4.3}{V}$ or $\SI{4.5}{V}$.}
\label{fig:fig5}
\end{figure*}

Using a model of NMC532 for the defective system and a open circuit voltage profile of NMC532 from Colclasure et al. \cite{colclasure2020electrode}, we perform reaction-diffusion simulations \change{with a single particle model as described in Section \ref{sec:rxn_diff}} to study the surface degradation of nickel-rich electrodes and are able to qualitatively reproduce the high voltage growth of the cation disordered phase at the surface using a single particle model to simulate cycling in an electrochemical cell.
A single particle model is able to capture the electrochemical behavior without the complexity of electrolyte diffusion limitation or cell size limitations.
In these simulations, we do not aim for a perfect fit of the model, but attempt to show that qualitatively correct behavior can be achieved using these ideas.
Using a cutoff voltage of 4.4 V vs. Li/Li$^+$ for oxygen formation, we can reproduce the behavior observed at surfaces for nickel rich electrodes \cite{xu2011identifying, yan2015evolution} based on the fact that the overpotential at high voltages will be positive in some regions, increasing the oxygen formation reaction amount as in Fig. \ref{fig:fig5}a.
\change{The experimental results that were used for comparison were selected based on criteria described in Appendix \ref{appdx:experimental}.}

For this set of simulations, the exchange current density for intercalation and reorganization energy are taken to be roughly $i_r^* = \SI{8}{A/m^2}$ for NMC532 and $\lambda = 3.78 k_BT$ for NMC111 since data for NMC532 was not freely available \cite{Zhang2022_CIET_preprint}.
\change{In our simulations,} roughly \SI{20}{nm} of the cation disorder growth in the high voltage model was achieved using an oxygen reaction parameter of $k_{0, oxy} = 5 \times10^{-6} \SI{}{A/m^2}$.
The diffusion coefficient for intercalation was assumed to be $D_c = 1 \times 10^{-12} (1-v) \SI{}{m^2/s}$ \cite{hong2020revealing, gao2018revealing} by the scale of diffusion measured experimentally, while for defects it was assumed to be $D_v = 5 \times 10^{-24} \SI{}{m^2/s}$.
A single spherical particle of radius $R = \SI{100}{nm}$ was used to model a NMC battery nanoparticle with a discretization of 200 finite difference volumes\change{, where details are seen in Appendix \ref{appdx:numerical}}.
This particle defect concentration initialized at a homogeneously distributed distribution of the optimal defect concentration of 2\% in the particle \cite{tang2019facilitating}, \change{so the initial concentration $v(t=0) = 0.02$}.
\change{The initial concentration of lithium in NMC532 $c(t=0)$ was set to 0.4.}

Single particles \change{of NMC532 such that $x=0.5$} are cycled at $\SI{1}{C}$ for 100 cycles, with a set of low voltage simulations with higher voltage cutoffs up to $\SI{4.3}{V}$ and a set of high voltage simulations up to $\SI{4.5}{V}$, modeling the experimental behavior in Yan et al. \cite{yan2017atomic}.
\change{The lower voltage cutoff was permanently set to $\SI{3.75}{V}$.}
\change{More information on the numerical simulations can be seen in Appendix \ref{appdx:numerical}.}
The amount of cation disorder growth post cycling and the overpotentials in the simulations are shown in Figs. \ref{fig:fig5}b and \ref{fig:fig5}c.
In these ranges, based on the asymmetry of the Butler-Volmer reaction, there is more oxygen formation at the surface of the particle, causing cation disorder to initiate at the surface and diffuse inward.
At high voltages, the amount of cation disorder nears the amount measured in experiments to be roughly \SI{20}{nm}, close to the experimental measurement of \SI{25}{nm} of disorder \cite{yan2017atomic}.
The surface phase appears when the amount of defects is high and grows inward towards the center of the particle as the amount increases.
Meanwhile, there is almost no growth for the low voltage phase, similarly to the experimental measurement of \SI{2}{nm} of cation disorder \cite{yan2017atomic}.
From our simulated results, we see that there is a larger amount of capacity fade happening after 100 cycles at higher voltages than at lower, with most of it happening at the surface of the particle at higher voltages. 
We see our model is able to reproduce the experimental data observed qualitatively.

\subsection{Voltage Hold}

\begin{figure*}[t]
\begin{center}
    \includegraphics[width=6.5in]{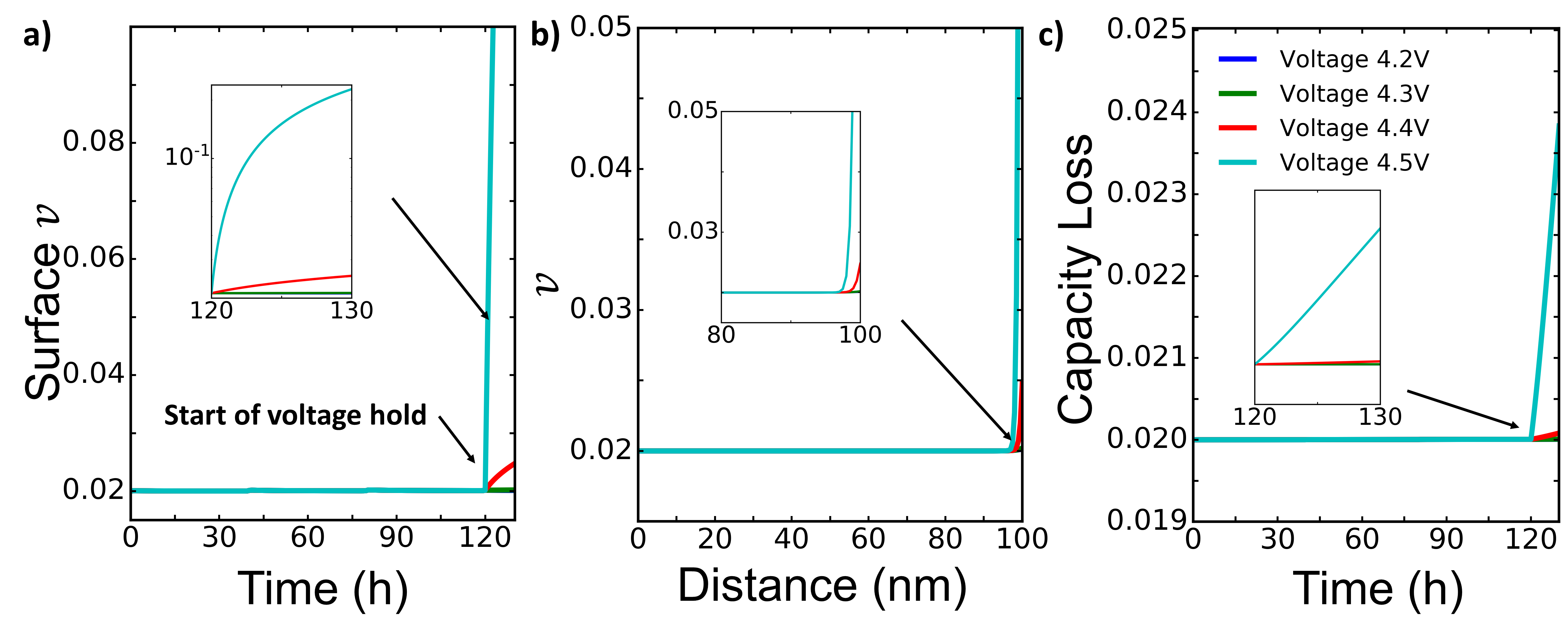}
\end{center}
\caption{Three C/20 cycles are performed before a \SI{10}{h} voltage hold is applied to particles at voltage of \SI{4.2}{V}, \SI{4.3}{V}, and \SI{4.4}{V} for NMC532. a) The surface defect amount $v$ is plotted as a function of the fraction of the total simulation time. The voltage hold occurs around \SI{60}{h} in the simulation. b) The variation of the capacity loss within a battery, where the distance is from the center of the particle to the edge, is plotted. c) The total capacity loss in the battery is plotted as a function of the fraction of total simulation time.}
\label{fig:fig7}
\end{figure*}

In addition to cycling results from a constant current perspective, the cutoff voltage with respect to the system is also an important parameter.
Voltage hold tests are also often performed to understand the degradation of batteries \cite{tornheim2019effect, mohanty2015understanding} at higher voltage.
Three constant current cycles at C/20 were performed and constant voltage holds of \SI{10}{h} were performed for the NMC532 single particle model described above in Fig. \ref{fig:fig7}.
The effect of the oxygen reaction potential, which is applied at \SI{4.4}{V} in our system, was found to be quite significant.
Large amounts of capacity growth are found at voltage holds past the oxygen reaction potential, while minimal amounts are found at lower voltages.
The amount of time spent at higher potentials is critical to controlling the amount of degradation in the particle, which is also observed experimentally \cite{jung2017oxygen}.

From the simulated data for constant current cycling and voltage holds at high potentials shown, we see that high voltage cycling in nickel rich electrodes causes irreversible effects on the degradation of the particle.
The degradation behavior shows that the operation time at lower voltages contributes negligibly to degradation of the electrode.
Thus, for cases where preventing degradation is extremely important, avoiding the higher voltage range is crucial.
If higher voltage operation is necessary, higher voltage operation should be applied later in the operation of the battery, to push back the onset of irreversible degradation.
\change{Experimental data from voltage hold simulations can be used to ``invert'' kinetic parameters for the degradation models described in these papers to infer more accurate exchange current densities for degradation reactions as shown in Appendix \ref{appdx:analytical_sol}.
Data provided at different cutoff voltages provides very impactful information on the voltage cutoffs for when degradation occurs \cite{csernica2021persistent}.}

The application of different coatings or additives may help reduce the amount of oxygen reaction in the particle, helping control the amount of degradation \cite{mauger2014surface}.
Commercialized batteries, especially for nickel rich materials, have different types of additives and coatings, such as aluminium oxides \cite{han2017coating}, carbon coatings \cite{gabrisch2006carbon}, and others \cite{baggetto2013surface}.
These can change the surface kinetics and dielectric properties of materials.
\change{Coating materials reduce the amount of degradation by reacting with the surface layers to form a more stable interface, changing the kinetic properties of degradation.
Using such related models to understand how coatings change the kinetic properties of degradation, or the redox potential at the interface, may prove extremely useful in future material design.}

\section{Conclusion}\label{sec:conclusion}
In this study, we show that we are able to formulate a degradation model of cation disorder coupled with oxygen formation at the surface from first principles for nickel-rich layered materials, which because of the high commercial availability of nickel, are slowly becoming the next generation of battery cathode materials.
These free energy models, combined with an oxygen vacancy boundary condition, are able to qualitatively explain why high voltage cycling causes more phase transformations and disorder \cite{yan2017atomic} in battery materials.
This is the first (to the authors' knowledge) model of free energy for cation degradation that can be easily derived from first principles which is applicable to continuum-scale battery simulations, combining the first-principles understanding of crystal structures with the computational tractability of battery models at the continuum level.
The applicability of these models is high, as no experimental data is needed to characterize the chemical potential functions or free energy models for degradation.
Experimental data is only required to characterize the kinetic parameters of degradation.

These types of degradation models can be applied to porous electrode models to study the effects of particle size, charging rate, and other macroscopic battery parameters on cation disorder formation in the active materials  \cite{fuller1994simulation, doyle1993modeling}.
Cycling simulations at different operating conditions, such as constant current and constant voltage, can be simulated to model their effect on the inducing of cation disorder and thus phase transformations in a battery, a crucial step to the study of long term maintenance of battery operation.
Insight from single particle cycling simulations of these particles reveal that the degradation caused by higher voltage operation in these materials is irreversible and that if avoiding degradation is imperative, higher voltage ranges should be applied later in the operation of a battery.
With respect to materials design, from a thermodynamic perspective, it is preferable to maintain a balanced Ni-Co-Mn ratio, since Ni-rich materials, despite their more attractive techno-economics, are predicted to be less stable with increasing Ni concentration, which promotes blocking of lithium sites by nickel disorder \cite{noh2013comparison}.

Our theory of degradation also may help to guide the development surface modifications to stabilize transition metal oxides.  
From a kinetic perspective, for a given NMC composition, the surface treatment must limit oxidation reactions that trigger cation disorder by passivating the reactive oxygen ions at the edge plane and blocking electron transfer to the electrolyte. Well known ceramic coatings, such as alumina (Al$_2$O$_3$), zirconia (ZrO$_2$), magnesium oxide (MgO), and other oxide materials, are able to perform these functions and can extend the cycle life of of nickel-rich oxides \cite{appapillai2007microstructure, han2017coating, zhang2020revealing}. Some coatings, such as niobium oxides (NbO), not only passivate oxygen and block electron transfer, but also introduce more stable ions, such as Nb$^{5+}$, into the crystal structure near the surface at transition metal or lithium sites ~\cite{xin2019li}. Our theory would suggest that such inserted ions from the coating may exchange with the more unstable nickel ions and reduce the tendency for cation reduction and disorder at the surface.

\change{
Degradation and synthesis are also related and can be considered inverse processes of each other \cite{das2017first}.
Thus, predictive capabilities of defects may also understand and guide direct synthesis of nickel-rich materials to reduce the number of defects during synthesis, which is generally a trial-and-error driven field \cite{bianchini2020situ}.
Understanding the equilibrium behavior of the materials could prevent the formation of these defects by synthesizing at specific concentrations, temperatures, or compositions of material that lead to reduced cation disorder.
This could be highly related to the temperatures and hold times used in the synthesis process.
}

Our general theoretical framework for cation disorder driven by electrochemical reactions may find applications in other uses of intercalation materials \cite{sood2021electrochemical}.  For example, \change{ transition metal oxides, such as lithium manganese oxide (LMO) and various NMCs,} have been used for electrochemical lithium extraction from aqueous multicomponent brines or seawater, \change{ but they suffer from the same problems of electrochemical stability as in Li-ion batteries~\cite{wu2022lithium}. Cation disorder can also play a role in limiting the performance of other Li-ion battery materials such as Li$_x$FePO$_4$ (LFP), which naturally contains a small fraction of iron anti-site defects that inhibit lithium diffusion~\cite{malik2010particle}. In electrochemical lithium extraction from brines,} even small amounts of Mn$^{2+}$ intercalation can irreversibly poison the material~\cite{zhao2013li} and large amounts of intercalated Na$^+$ interfere with selectivity for Li$^+$ ~\cite{liu2020lithium}. \change{ As suggested by our theory, experiments have shown that it can be helpful to avoid strongly oxidizing conditions at high potentials to preserve the crystal structure~\cite{zhao2013li,guo2020effect}}. More generally, electrochemical methods of water treatment and recovery of critical minerals based on electrosorption using intercalation electrodes often suffer from anti-site defects that block available sites, thereby lowering ion removal rates and storage capacity~\cite{alkhadra2022_electrochemical_methods,srimuk2020charge}. There is currently no available theory of competitive adsorption and cation-defect formation to model these important phenomena, but our theory may provide a natural starting point.

\bigskip

\begin{acknowledgments}
This work was supported by the Toyota Research Institute through D3BATT: Center for Data-Driven Design of Li-Ion Batteries.
The authors would especially like to acknowledge Dr. Dimitrios Fraggedakis, Dr. Huada Lian, Huanhuan Tian, Dr. J. Pedro de Souza, Dr. Junghwa Lee, and Prof. Mehran Kardar for valuable discussions and insight.
\end{acknowledgments}
\include{acknowledgements}

\appendix 

\section{Diffusional Chemical Potentials} \label{appdx:chem_pot}

Simple formulae for the entropic contributions to the diffusional chemical potentials for intercalation and defects, respectively, are given by
\begin{eqnarray}
    -T \frac{\partial S}{\partial c} &=& 
    k_BT \ln{\frac{c}{1-c-v}}\\
    -T \frac{\partial S}{\partial v} &=&
    k_BT \left( \ln{\frac{v}{x-v}} +\ln{\frac{v}{1-c-v}}\right)
\end{eqnarray}
based on our ideal solid solution models for lithium ion intercalation between the nickel oxide layers and for vacancy-mediated defects within the layer. Expressions for the enthalpic contributions from dipole-dipole interactions, core energy, and dielectric decrement are more complicated and must be calculated numerically. 
Combining the enthalpic and entropic contributions, we obtain the total diffusional chemical potentials for lithium intercalation and vacancy-mediated defects, respectively:

\begin{eqnarray}
\mu_{c} &=& \frac{\partial H}{\partial c} \\
&=& \nonumber
- \frac{qe}{4\pi \varepsilon} \left( \sum_{i \in \text{even}} \sum_{j \in i} \frac{\boldsymbol{\mu}_{ij}\cdot \mathbf{r}_{ij,0}}{|r_{ij,0}|^3} -  \sum_{i \in \text{odd}} \sum_{j \in i} \frac{\boldsymbol{\mu}_{ij}\cdot \mathbf{r}_{ij,0}}{|r_{ij,0}|^3}\right) + \frac{(qe)^2(1-v)}{12\pi\varepsilon r_0} -\frac{\Delta H(c, v)}{\varepsilon}\frac{\partial \varepsilon}{\partial c}\\
&+& k_BT \ln{\frac{c}{1-c-v}} \\
    \mu_{v} &=& \frac{\partial H}{\partial v} \\
&=& \nonumber\frac{cqe^2 \text{EN}_{Ni}} {2\pi \varepsilon} \left( \sum_{i \in \text{even}} \sum_{j \in i} \frac{\boldsymbol{\mu}_{ij}\cdot \mathbf{r}_{ij,0}}{\mu_{ij}|r_{ij,0}|^3} -  \sum_{i \in \text{odd}} \sum_{j \in i} \frac{\boldsymbol{\mu}_{ij}\cdot \mathbf{r}_{ij,0}}{\mu_{ij}|r_{ij,0}|^3}\right) - \frac{(qe)^2 c}{12\pi\varepsilon r_0}-\frac{\Delta H(c, v)}{\varepsilon}\frac{\partial \varepsilon}{\partial v} \\
 &+& k_BT \left( \ln{\frac{v}{x-v}} +\ln{\frac{v}{1-c-v}}\right)
\end{eqnarray}

where $\mu_{ij} = |\boldsymbol{\mu}_{ij}|$.

\section{Convergence of Calculations} \label{appdx:convergence}

\begin{figure}[t]
\begin{center}
\includegraphics[width=2.5in]{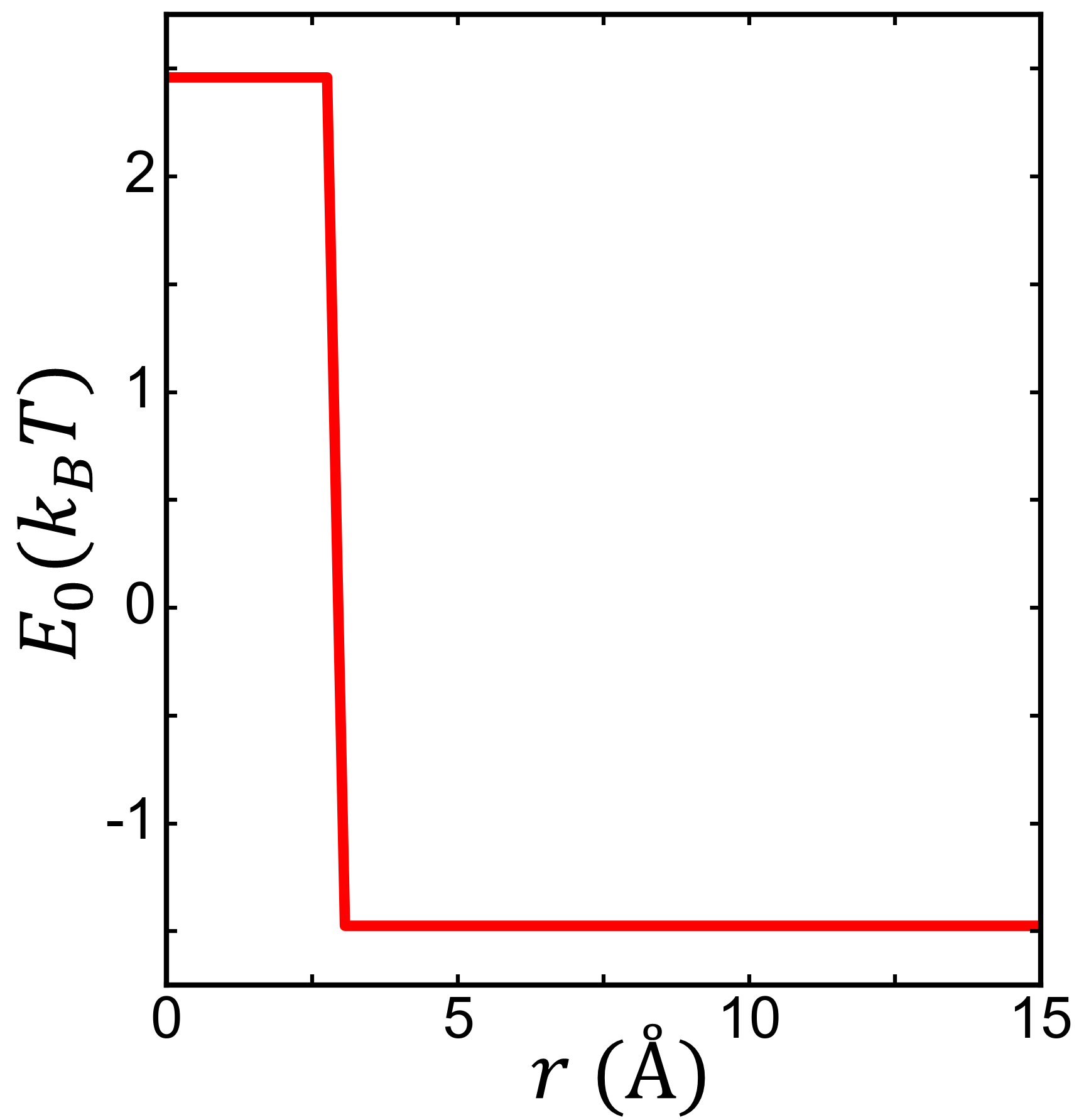}
\end{center}
\caption{Convergence of electrostatic calculations with respect to the cutoff distance from the center of the defect.}
\label{fig:appendix1}
\end{figure}

The convergence of electrostatic dipole-charge calculations is shown in Fig. \ref{fig:appendix1}.
It is well known that charge-charge interactions in an electrostatic system will not converge above three dimensions \cite{kardar2007statistical}.
However, based on the order of magnitude reduction from dipole-charge interactions, these interactions converge quite quickly.

\section{Dielectric Constant Calculations}
The dielectric constant plotted with the Maxwell-Garnett equation for Li$_c$Ni$_x$Mn$_y$Co$_z$O$_2$ is shown as below as the additive rule.
The additive rule is 
\begin{equation}
    \frac{\varepsilon-1}{\varepsilon+2} = \frac{4}{3}\pi\sum_{i}\alpha_i n_i,
\end{equation}
where $\alpha_i$ is the polarizability of the atom $i$ and $n_i$ is the number density of atom $i$ \cite{shannon1993dielectric, coker1976empirical}.
There is an decrease the dielectric constant with the amount of decrease in lithium concentration in the material.

\begin{figure}[t]
\includegraphics[width=0.5\textwidth]{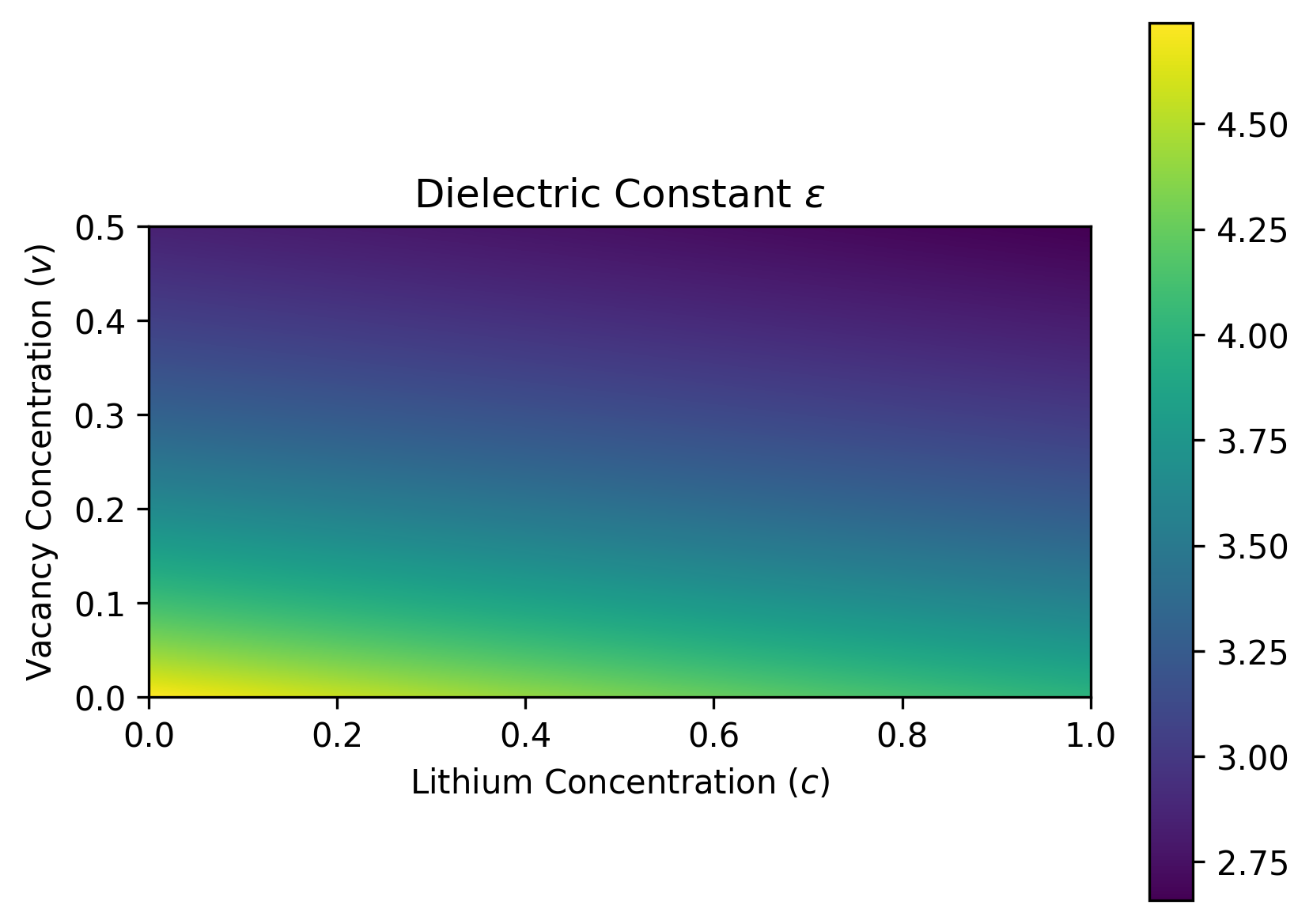}
\centering
\caption{Dielectric Constant Measured with the Additive Rule.}
\label{fig:appendix3}
\end{figure}

\section{Analytical Solution} \label{appdx:analytical_sol}

Assuming solid diffusion is not limiting, using a simplified single particle model, we can predict the scaling of the amount of degradation in the first few cycles with a voltage hold.
Here, we first start with the reaction rate integrated over time
\begin{equation}
\begin{split}
    Q_{loss} & = \int_0^\tau j_{deg} A dt \\
    & = \int_{0}^1 \tau k_{0,deg} \exp{(\eta + \mu(c) - \mu_{deg})} dc \\
    & \propto \tau\int_0^1 \exp{(2\text{arcsinh}\left(\frac{R c_{max}}{6 \tau F k_0(c)} \right) + \mu(c) - \mu_{deg})} dc \\ 
    & \propto 
    \tau N \int_0^1 \exp{(2\text{arcsinh}\left(\frac{R c_{max}}{6 \tau F k_0(c)} \right) + \mu(c) - \mu_{deg})} dc.
    \end{split} 
\end{equation}
Since the exponential term is quite small when the voltage is not above the cutoff voltage, the capacity loss scales with the amount of time spent in a higher voltage regime, which indicates a linear scaling with the number of cycles.
For a voltage hold, this equation simplifies to
\begin{equation}
\begin{split}
    Q_{loss} & =
    \int_0^\tau k_{0,deg}\exp{(V-\mu_{deg})}dt \\
    & = k_{0,deg} \tau \exp{(V - \mu_{deg})}.
    \end{split}
\end{equation}
This relation is linear with respect to time spent in the voltage hold, as shown in Fig. \ref{fig:fig7}c, and depends exponentially on the value of the voltage hold.
If we assume the degradation voltage is at $\mu_{deg}$, then the exchange current density for voltage loss is generally
\begin{equation}
    k_{0,deg} = \frac{Q_{loss}}{\tau \exp{(V - \mu_{deg})}}.
\end{equation}

\section{Numerical Implementation} \label{appdx:numerical}
The two separate numerical calculations were both implemented in MATLAB and can be found at \url{https://github.com/lightningclaw001/public_paper_scripts/tree/main/cation_disorder_defect}.
The thermodynamic model was calculated using the unit cell of Materials Project structure mp-632864 for LiNiO$_2$ \cite{jain2013a, jain2013b, laubach2009, hirano1995, adipranoto2014}.
The unit cell was reproduced and based on the convergence calculations in Appendix \ref{appdx:convergence}, a cutoff of 20 \AA.
Atoms within the convergence criteria were summed with Eq. \ref{eq:enthalpy} to find the enthalpy, which combined with Eq. \ref{eq:entropy}, gives the total free energy $G = H - TS$.
The analytical solutions to the chemical potentials were calculated as Appendix \ref{appdx:chem_pot}, or as $\mu_c = \frac{\delta G}{\delta c}$ or $\mu_v = \frac{\delta G}{\delta v}$.

A spherical single particle finite difference model \cite{guo2010single} (reaction diffusion model) was implemented for Eqs. \ref{eq:cons_c}, \ref{eq:BC_c} for concentration, and Eqs. \ref{eq:cons_v}, \ref{eq:BC_v} for vacancies.
The sphere was discretized into $n$ sections, with $\Delta x$ as the size of each section, where the boundary conditions Eqs. \ref{eq:BC_c} and \ref{eq:BC_v} were applied at the edges, and $r_i$ is the radius at the center of each discretization.
The fluxes $F_{i-0.5}$ were defined at the edges of the discretizations (at $r_{i-0.5}$).
The bulk equations for concentration were then \begin{equation}
    \frac{dc_i}{dt} = -\frac{1}{r_i^2} \frac{(r_{i+0.5}^2 F_{c,i+0.5} - r_{i-0.5}^2 F_{c,i-0.5}) }{\Delta x},
\end{equation}
where the fluxes are defined as
\begin{equation}
    F_{c,i+0.5} = - \frac{(D_c c (1-v) \mu_c) \big \lvert_{i+1}-(D_c c (1-v) \mu_c) \big \lvert_{i}}{\Delta x}.
\end{equation}
At the boundary, the flux $F_{c,n+0.5} = -i_{int}$ is related to the intercalation current.

For the vacancies in the system, we have similar discretizations such that
\begin{equation}
    \frac{dv_i}{dt} = -\frac{1}{r_i^2} \frac{(r_{i+0.5}^2 F_{v,i+0.5} - r_{i-0.5}^2 F_{v,i-0.5}) }{\Delta x},
\end{equation}
where the fluxes are defined as
\begin{equation}
    F_{v,i+0.5} = - \frac{(D_v v \mu_v) \big \lvert_{i+1}-(D_v v \mu_v) \big \lvert_{i}}{\Delta x}.
\end{equation}
At the boundary, the flux $F_{v,n+0.5} = -i_{v} = 2 i_{oxy}$ is related to the oxygen degradation current, where the conservation equation is written as
\begin{equation}
    \frac{\partial c_{oxy}}{\partial t} =  2 \pi r_n \Delta x ~i_{oxy}.
\end{equation}

An additional algebraic constraint for the current constraint was added so that 
\begin{equation}
    \sum_{i=1}^n v_i \frac{dc_i}{dt} = R_{constraint}, 
\end{equation}
where $v_n$ is the volume fraction of each section, $v_i = \frac{r_{i+0.5}^3 - r_{i-0.5}^3}{r_n^3}$.
The ode solver ode15s was used to solve this problem with relative and absolute tolerances of $1\times 10^{-8}$.

\section{Symbols} \label{appdx:symbols}
Here, we have appended a table of the symbols used in this paper for ease of understanding:
\begin{widetext}
\begin{center}
\begin{tabular}{ |c|c|c| } 
 \hline
 Symbol & Meaning & Units \\ 
 \hline
 $c$ & lithium concentration in the solid material & dimensionless \\ 
 $c_+$ & Li$^+$ concentration in the electrolyte & M \\ 
 $v$ & defect concentration & dimensionless \\ 
 $x$ & nickel concentration in NMC material & dimensionless \\
 $y$ & manganese concentration in NMC material & dimensionless \\ 
 $z$ & cobalt concentration in NMC material & dimensionless \\
 $\boldsymbol{\mu}_{0}$ & dipole pointing from transition metal layer to the lithium layer & C$\cdot$m \\
 $\mathbf{r}_{ij}$ & distance vector between the defect center and site $ij$ & m \\
 $r_{ij}$ & magnitude of distance vector between the defect center and site $ij$ & m \\
$ \varepsilon$ & dielectric constant, see Appendix C & F/m\\
 $\mathbf{r}_0$ & distance vector between a transition metal-lithium dipole & m \\
 $r_0$ & magnitude of distance vector between a transition metal-lithium dipole & m \\
 $H$ & enthalpy & $eV$ \\
 $S$ & entropy & $eV/K$ \\
 $G$ & Gibbs free energy, $G = H -TS$ & $eV$ \\
 $\mu_c$ & chemical potential for intercalation & $k_B T$ \\
 $\mu_v$ & chemical potential for defect formation & $k_B T$ \\
 $D_c$ & diffusion coefficient for lithium concentration & m$^2$/s \\
 $\mathbf{F}_c$ & flux for lithium diffusion & m/s \\
 $D_v$ & diffusion coefficient for defect formation & m$^2$/s \\
 $\mathbf{F}_v$ & flux for defect formation & m/s \\
 $i_{int}$ & intercalation current & A/m$^2$ s \\
 $i_{oxy}$ & oxygen degradation current & A/m$^2$ s \\
 $i_r^*$ & exchange current density & A/m$^2$ s\\
 $\lambda$ & reorganization energy & $k_BT$ \\
 $\eta$ & overpotential & $k_B T/e$ \\
 $\eta_f$ & formal overpotential, $\eta_f = \eta + \ln{\frac{c_+}{c}}$ & $k_B T/e$ \\
 $\phi$ & potential difference between electrolyte and solid & $k_B T$ \\
 \hline
\end{tabular}
\end{center}
\end{widetext}

\section{Experimental Data Selection} \label{appdx:experimental}
\change{
To obtain experimental data about disorder, recent microscopy experiments are very useful in discovering the amount of disorder in the system. 
Many experimentalists use gas formation or other mechanisms \cite{jung2017oxygen} to deduce the amount of phase transitions, but for our model specifically studying cation disorder, it is imperative to have a spatially defined set of experiments with clear experimental cutoffs for a specific material. 
In addition, many common experimental measurements for phase transitions also measure the oxidation state of different atoms, but this does not give the necessary information because it discusses more the specific phases that we see \cite{lin2014surface}.
Since our paper is focused on cation disorder, which is the trigger for phase transformations, knowing the amount of densified phases such as rock salt or spinel phases does not provide the exact data we want. 
We have chosen experimental data with a good set of electrochemical range (voltage cutoffs) as well as spatial range, along with a material that we are interested in, which needs to be nickel rich and layered \cite{yan2017atomic}. 
Similar material is also well-characterized for modeling purposes in terms of open circuit voltage and electrochemical parameters \cite{colclasure2020electrode}, which was also previously verified and tested by the authors, and compared with other experiments and simulations \cite{park2021fictitious}. 
Coupled ion electron transfer kinetics parameters for nickel rich materials were also obtained with this set of materials \cite{Zhang2022_CIET_preprint}. 
Thus, we chose this set of experiments to model and compare because of a) rigor of experiments and well-planned and defined experiments b) availability of modeling infrastructure and kinetic and thermodynamic parameters.
}

\bibliography{biblio}
\bibliographystyle{abbrv}

\end{document}